\documentclass[superscriptaddress,showpacs,aps,twocolumn,prb,floatfix]{revtex4-2} 
\usepackage{eurosym}
\usepackage{amssymb}
\usepackage{amsmath}
\usepackage{amsfonts}
\usepackage{amsbsy}
\usepackage{graphicx}
\usepackage{bm}
\usepackage{xcolor}
\usepackage{hyperref}
\usepackage{units}
\usepackage{upgreek}
\usepackage{textcomp}

\usepackage{verbatim}

\usepackage[normalem]{ulem}

\begin{document}

\title{Influence of oxygen on electronic correlation and transport in iron in the outer Earth's core} 
  \author{G. G. Blesio}
\affiliation{Jo\v{z}ef  Stefan  Institute,  Jamova  39,  SI-1000  Ljubljana,  Slovenia}
\email{german.blesio@ijs.si}

\author{L. V. Pourovskii}
\affiliation{CPHT, CNRS, Ecole polytechnique, Institut Polytechnique de Paris, Route de Saclay, 91128 Palaiseau, France.}
\affiliation{Coll\`{e}ge de France, 11 place Marcelin Berthelot, 75005 Paris, France.}

\author{M. Aichhorn}
\affiliation{Institute of Theoretical and Computational Physics, Graz University of Technology, NAWI Graz, Petersgasse 16, Graz, 8010, Austria}
\author{M. Pozzo} 
\affiliation{Department of Earth Sciences and London Centre for Nanotechnology, University College London, Gower Street, London WC1E 6BT, UK}
\author{D. Alf\`{e}} 
\affiliation{Department of Earth Sciences and London Centre for Nanotechnology, University College London, Gower Street, London WC1E 6BT, UK}
\affiliation{Dipartimento di Fisica Ettore Pancini, Universit\`{a} di Napoli Federico II, Monte S. Angelo, I-80126 Napoli, Italy.}

\author{J. Mravlje}
\affiliation{Jo\v{z}ef  Stefan  Institute,  Jamova  39,  SI-1000  Ljubljana,  Slovenia}

\begin{abstract}
Knowing the transport properties of iron under realistic conditions
present in the Earth's core is essential for the geophysical modeling
of Earth's magnetic field generation. Besides by extreme pressures and
temperatures, transport may be influenced importantly also by the
presence of light elements. Using a combination of molecular dynamics,
density functional theory, and dynamical mean-field theory methods we
investigate how oxygen impurities influence the electronic
correlations and transport in the liquid outer Earth's core.  We
consider a case with an oxygen content of $\sim 10 \mathrm{atomic} \%$, a
value that is believed to be close to the composition of the core.
We find that the electronic correlations are enhanced but their effect on conductivities is moderate (compared to  pure Fe, electrical conductivity drops by 10\% and thermal conductivity by 18\%). 
 The effect of electron-electron scattering
alone, whereas not large, is comparable to effects of the
compositional disorder.  We reveal the mechanism behind the larger
suppression of the thermal conductivity and associated reduction of
the Lorenz ratio and discuss its geophysical significance.

\end{abstract}
 
\maketitle
\section*{Introduction}
\label{sec_introduction}

The Earth's magnetic field is generated by  a self-excited dynamo that is driven by
convection in the outer core.  The thermal conductivity of iron, which
determines the amount of heat flow available for convection, and the value of electrical conductivity, which determines the dissipation of the current, are 
crucial inputs for geophysical models of this geodynamo mechanism. Direct measurements of transport under
extreme pressures and temperatures that are relevant  to Earth's core are
challenging~\cite{Ohta2016,Konopkova2016} as one must ensure homogeneous
temperature and carefully control the geometry of the samples~\cite{Lobanov2022}.

One can access transport properties also from 
first principle calculations based on the molecular dynamics
(MD)--density functional theory (DFT)
method~\cite{Alfe1998,Pozzo2012,deKoker2012}.  These calculations have
shown that the electrical and thermal
conductivities have values that are significantly (2-3 times) higher~\cite{Pozzo2012,deKoker2012} than 
earlier established estimates~\cite{Stacey2001,Stacey2007}. These earlier estimates were 
based on extrapolations that neglected the effects of resistivity
saturation~\cite{Gunnarsson2003,Gomi2013,Gomi2016}.  The higher values of thermal
conductivity lead to a different geophysical picture, with an
inner core  that is younger ($<1$  billion years, whereas   magnetism is known to exist for at least $3.4$ billion years~\cite{Tarduno2010,Tarduno2015}
), and less thermal
convective energy to drive the geodynamo, which is known as ``the new core
paradox''~\cite{Olson2013}. Namely, less thermal energy implies
that convection must be helped by the chemical convection driven by
the exsolution of lighter elements but this was less active before the formation of the
inner core.

There has been an ongoing discussion on whether electronic correlations, which have been shown to be significant in iron under Earth's core conditions~\cite{pourovskii2013,Pourovskii_2019}, can cause a breakdown in the Mott-Ioffe-Regel resistivity saturation~\cite{Gunnarsson2003,Deng2013}. This debate revolves around whether these correlations are strong enough to reduce conductivity values to previously established levels~\cite{Zhang2015,Pourovskii2017,Xu2018}, as suggested in a pioneering work (later retracted)\cite{Zhang2015}. Recent findings indicate that electron-electron scattering (EES) plays only a moderate role and represents a small fraction compared to thermal-disorder (electron-phonon) scattering~\cite{Pourovskii2020,Zhang2022}.

One important question remains to be addressed: does alloying with lighter elements significantly enhance correlations, and if so, to what extent are conductivities suppressed? The Earth's core contains a sizable contribution of lighter elements, primarily silicon and oxygen, with recent research indicating a high oxygen concentration~\cite{Badro2015}. The influence of these substitutions has been broadly investigated using MD-DFT~\cite{Pozzo2012,deKoker2012,Wagle2019,Li2021}, but without accounting for correlation effects. Oxygen might enhance electronic correlations by reducing iron 3d shell occupancy toward half-filling~\cite{Georges2013}. Indeed, recent theoretical work~\cite{Jang2021} investigated several ordered FeO structures and found a very strong enhancement of electron-electron scattering (EES). However, the impact of thermal disorder on EES was neglected in that study, and the final effect on thermal conductivity, taking into account both EES and electron-phonon scattering, was not evaluated. Furthermore, the impact of EES on conductivity in liquid iron, which is most relevant for the dynamo mechanism, was not clarified.

In this study, we investigate the transport properties of Fe and FeO alloys in their liquid state at the inner-core boundary (ICB) and core-mantle boundary (CMB) conditions. We describe the liquid state using MD-DFT and account for electron-electron scattering (EES) using dynamical mean-field theory (DMFT)~\cite{georges_dmft_1996,Hausoel2017,Xu2018,Pourovskii2020}. Specifically, we focus on the Fe$_{0.91}$O$_{0.09}$ 
composition and observe that its EES rate increases by $\sim 25\%$ compared to pure iron (Fe
), but the conductivities are affected to a lesser extent with only an $\approx 10\%$ drop in electrical conductivity $\sigma$ and an $\approx18\%$ drop in thermal conductivity $\kappa$. This change is mainly due to band-structure effects rather than a direct increase in EES.

In order to quantify the effects of electronic-correlation on
transport we  compare the calculated DMFT conductivities with those from
the MD-DFT. We find that the inclusion of EES leads to only a moderate reduction of
electrical and thermal conductivities by roughly 10\% and 20\%,
respectively. This finding is crucial considering a large body of existing theoretical work that neglects EES~\cite{Pozzo2012,deKoker2012,Gomi2016,Wagle2019}.
A related study of Fe-Si alloys has shown similar
behavior~\cite{Zhang2022}. We discuss why the thermal conductivities are
suppressed more than the electrical ones and highlight the geophysical
implications of our results.

\section*{Methods}

We performed the molecular dynamics calculations using the VASP
code~\cite{kresse1996}, using the projected augmented wave
method~\cite{kresse1999,blochl1994} to describe the interactions
between the electrons and the ions and expanded the single-particle
orbitals as linear combinations of plane waves (PW), including PW with
maximum energies of 400 eV. The molecular dynamics simulations were
performed by sampling the Brillouin zone using the $\Gamma$ point
only, and a time step was 1 fs. The temperature was controlled using
the Nos{\'{e}}~\cite{Nose1984} thermostat. To compute the DFT
electrical and thermal conductivity we used the modified version of
VASP by Dejarlais~\cite{Desjarlais2002}.

The DFT+DMFT self-consistent calculations were performed with a local
density approximation (LDA) approach in the Wien2K
code~\cite{Wien2K,Blaha2020}, using the TRIQS
library~\cite{triqs,triqs_wien2k_interface,triqs_wien2k_full_charge_SC,TRIQS/DFTTools}
for the DMFT and transport calculations. We used a local density-density interaction vertex with
interaction parameters $U = 5.0$\,eV, $J_H = 0.93$\,eV, in agreement with the previous studies on pure iron~\cite{Pourovskii2017,Pourovskii2020}, and solved the
impurity problem using the continuous-time hybridization-expansion
segment solver~\cite{ctseg1,ctseg2}. Each
calculation was first converged by 25 fully self-consistent DFT+DMFT
iterations, where each Monte Carlo run employed $2\times10^{10}$
Monte Carlo moves and 200 moves/measurement. Using the converged
Kohn–Sham Hamiltonian, 10 additional DMFT cycles were performed with
the number of Monte Carlo moves increased to $10^{11}$. To
obtain clean data for analytical continuation that we performed using Maximum Entropy method, 20 additional runs (with 
$2\times10^{11}$ moves per run) were
carried out starting from the same converged value of the DMFT bath
Green’s function and resetting the random sequence.

We calculated the conductivities within the Kubo linear-response neglecting the vertex corrections. The electrical and thermal conductivity read \cite{TRIQS/DFTTools,kotliar_elec_struc_dmft_2006}
\begin{equation}
\sigma_{\alpha\alpha'}=\frac{e^2}{k_B T} K^0_{\alpha\alpha'}\, ~, ~ \kappa_{\alpha\alpha'}=k_B\left[ K^2_{\alpha\alpha'} - \frac{\left(K^1_{\alpha\alpha'}\right)^2}{K^0_{\alpha\alpha'}} \right],
\label{eq:cond_w0}
\end{equation}
where $\alpha$ is the direction ($x$, $y$ or $z$) and $k_B$ the Boltzmann constant.
The kinetic coefficients $K^n_{\alpha\alpha'}$ are
\begin{equation}
K^n_{\alpha\alpha'}=2\pi\hbar\int d\omega (\beta\omega)^n f(\omega)f(-\omega)\Gamma^{\alpha\alpha'}(\omega,\omega),
\end{equation}
where 2 is the spin factor, $f(\omega)$ is the Fermi function, and the $\Gamma^{\alpha\alpha'}$ is given by
\begin{equation}
\Gamma^{\alpha\alpha'}(\omega,\omega')=\frac{1}{V}\sum\limits_{\mathbf{k}} \mathrm{Tr}\left( v^\alpha_\mathbf{k} A_\mathbf{k}(\omega) v^{\alpha'}_\mathbf{k} A_\mathbf{k}(\omega') \right)
\end{equation}
where $V$ is the unit-cell volume, $A_\mathbf{k}(\omega)$ is the DMFT spectral function at momentum $\mathbf{k}$, and $v^\alpha_\mathbf{k}$ is the corresponding band velocity in the direction $\alpha$.
We also define the  transport distribution 
\begin{equation} \Gamma(\omega) =\sum_\alpha \Gamma^{\alpha \alpha} (\omega, \omega).  \end{equation}

We also calculated the response at finite frequency $\Omega$ which yields the optical electrical and thermal  conductivity. These are evaluated by Eq.\ref{eq:cond_w0} using the kinetic coefficients evaluated at a finite frequency $\Omega$ 
\begin{align}
K^n_{\alpha\alpha'}(\Omega)=2\pi\hbar&\int d\omega \Gamma^{\alpha\alpha'}(\omega+\Omega/2,\omega-\Omega/2)  (\omega+\Omega/2)^n \notag\\
&\beta^{n-1}\frac{f(\omega-\Omega/2)-f(\omega+\Omega/2)}{\Omega}.
\end{align}

In the momentum sums we retained 14 momentum points. The electron-phonon-only values were calculated using the Kubo-Greeenwood approximation as implemented in VASP~\cite{Desjarlais2002} using 10 momentum points. We checked that upon increasing the number of momentum points further the results vary by less than 1\%. 

\section*{Results}
We calculate the liquid phase  for 67 atoms: all
iron (Fe
) or with oxygen (Fe$_{0.91}$O$_{0.09}$
). We study the liquid at CMB conditions for a
temperature $T=4400$ K, volume  $8.64\,$\AA$^3$/atom (which corresponds to  pressure 132 GPa), and ICB conditions $T=6350$ K with volume $7.16\,$\AA$^3$/atom (pressure 330 GPa).  

The calculations were performed for three snapshots separated by 5 ps time each and the calculations of transport properties was performed by averaging over the snapshots and spatial directions.

\begin{figure}[ht]
\begin{center}
\includegraphics*[width=\columnwidth]{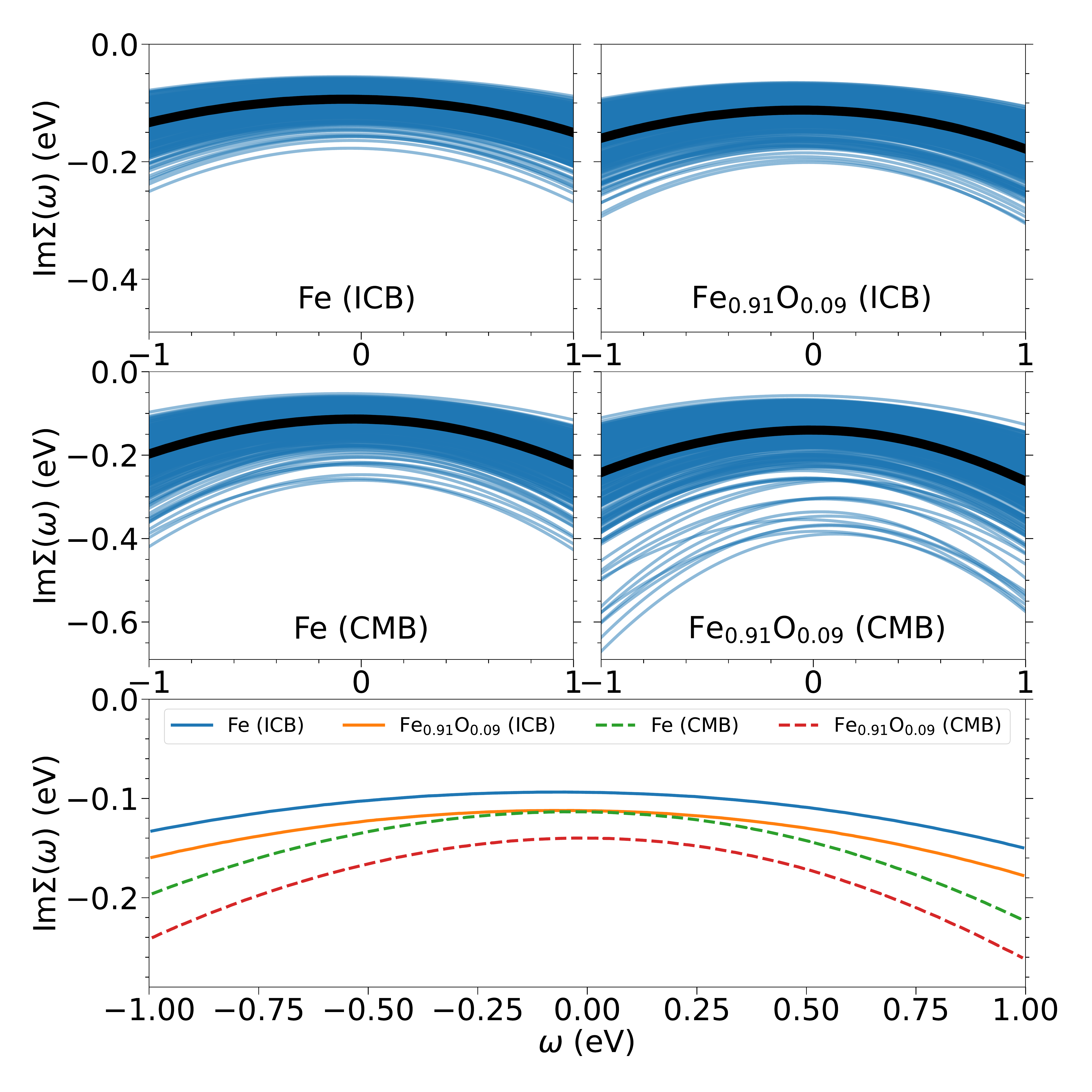}
\caption{Imaginary part of the calculated  self-energies on the real axis. In the top and middle panels, we show the individual
  self-energies (blue lines) and their average $\langle \mathrm{Im} \Sigma
  \rangle$ (black thick). In the bottom panel, the four average
  self-energies are shown, at the ICB 
  (full)
 and CMB 
  (dashed)  conditions.
}
\label{fig:self_w}
\end{center}
\end{figure}

{\bf Electronic correlations.} Figure~\ref{fig:self_w} displays the imaginary part of self-energies for a single snapshot in the energy range of $[-1,1]$ eV. Each curve represents a different site/orbital/spin index, and the thick lines indicate an average over all of them. Since the electron-electron scattering rate is $1/\tau= -2\mathrm{Im} \Sigma(\omega \rightarrow 0)$, more negative values of $\mathrm{Im} \Sigma$ indicate stronger electronic correlations. Our results show that the scattering increases significantly in the oxygen-rich case for both ICB and CMB. The distribution of the curves reveals that not only the Fe sites closest to oxygen are affected, but also the spread of the data is wider in the oxygen-rich case, and even the self-energies with the smallest magnitudes are enhanced by oxygen. Overall, the average EES, as given by the average $\mathrm{Im} \Sigma$, increases by 25\%.

\begin{figure}[ht]
\begin{center}
  \includegraphics*[width=\columnwidth]{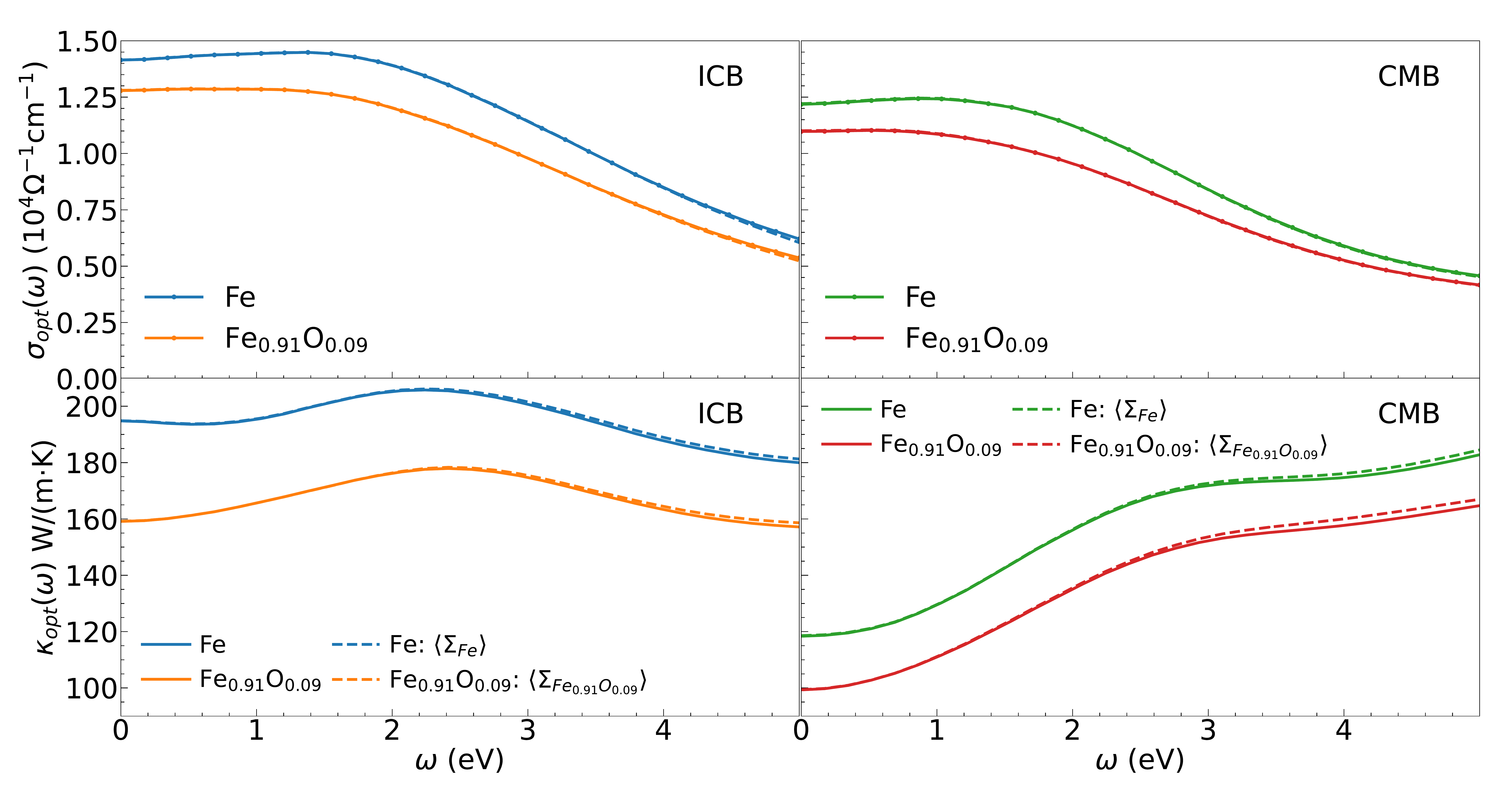}
  \caption{Optical electrical (top) and thermal (bottom) conductivity
    for pure Fe 
     and Fe$_{0.91}$O$_{0.09}$
     . The ICB 
    and CMB cases
    are shown on the left and right, respectively. The dashed
    lines (that overlap closely with the filled ones) indicate a
    simplified calculation using fully (site, orbital, and spin)
    averaged self-energies.  }
\label{fig:cond}
\end{center}
\end{figure}

{\bf Conductivities.} The calculated optical conductivities are shown in Fig.~\ref{fig:cond} for the case of electrical and thermal current on the top and bottom panels, respectively. The $\omega \rightarrow 0$ values indicate the dc-transport values. One sees that quantitatively the effect of oxygen on transport is somewhat weaker than on the EES. Also shown in that figure are the results of a simplified calculation where one uses the $\langle \Sigma \rangle$ instead of the individual self-energies. One sees a ``self-averaging'' effect: the results of such a calculation are almost indistinguishable from the full calculation. This also tells that statistical uncertainties of the individual self-energies will not affect the calculated conductivities.

\begin{figure}[ht]
\begin{center}
\includegraphics*[width=0.8\columnwidth]{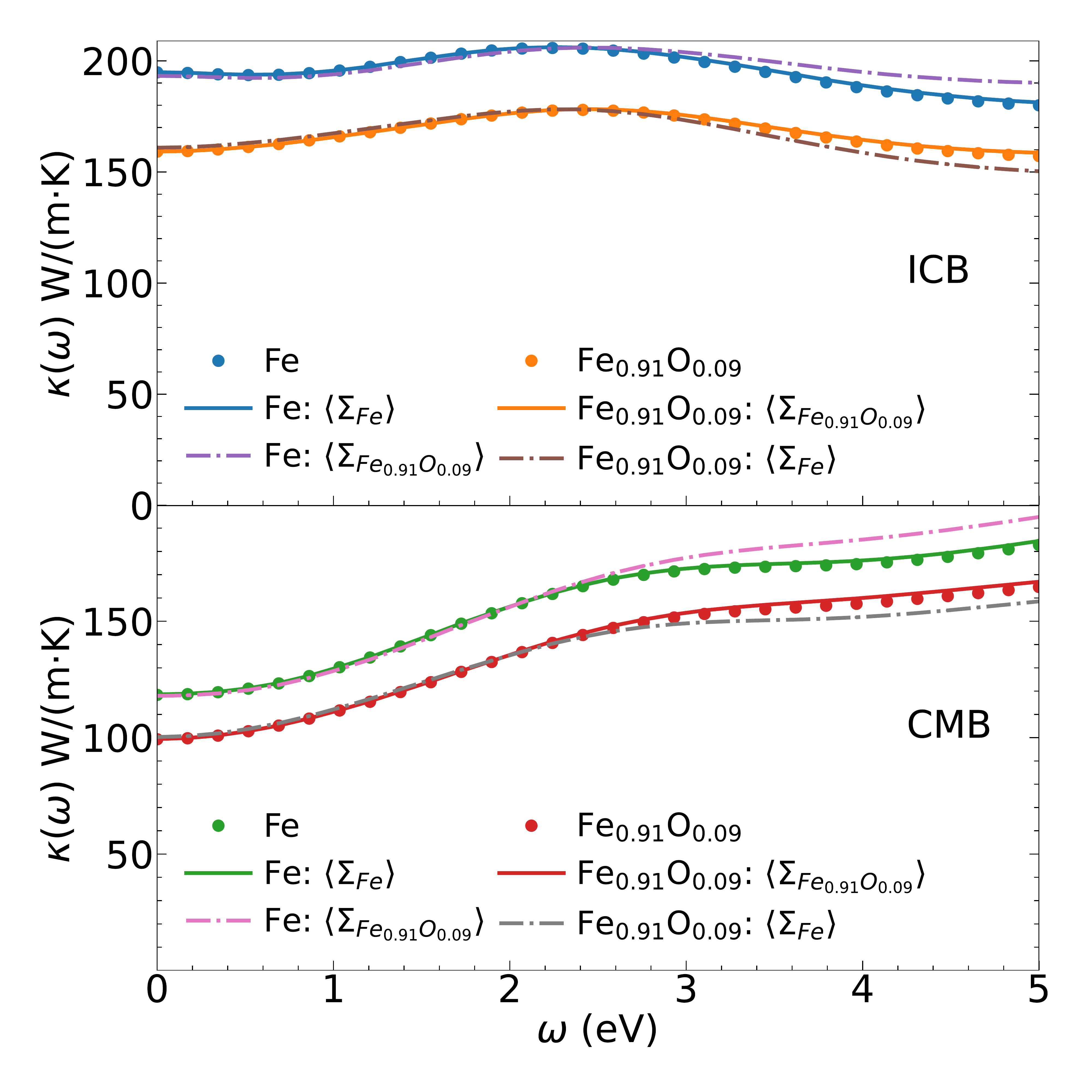}
\caption{Thermal conductivity for pure Fe 
 and Fe$_{0.91}$O$_{0.09}$ 
 for the ICB (top) and CMB cases (bottom). 
We show the results obtained with orbitally and site-resolved self-energies (dots) as well as those calculated using the  average self-energy (full line). The latter was first averaged over all sites and orbitals on the Matsubara grid and then analytically continued.
 The differences between the two are very small. With dash-dotted line, we show results  calculated by exchanging the average self-energy between the  Fe and Fe$_{0.91}$O$_{0.09}$. }
\label{fig:cond_exch}
\end{center}
\end{figure}

To what extent is the suppression of conductivities in the oxygen-rich case due to the increase of EES documented in Fig.~\ref{fig:self_w}?  It turns out that, as suggested in the earlier work~\cite{Pourovskii2020} iron under Earth's core conditions is in a thermal-disorder dominated case where the changes of EES impact transport weakly. In Fig.~\ref{fig:cond_exch} we demonstrate this by additional calculations where we compute the conductivity of the Fe$_{0.91}$O$_{0.09}$ case by using the scattering information from $\langle \Sigma \rangle$ corresponding to pure Fe calculation and vice versa for the other case. Quite strikingly, these ``exchanged'' calculations are at small frequencies almost indistinguishable from  the ``non-exchanged'' ones. This tells that the oxygen affects the results through a structurally induced change in the band dispersions and that the changes in the EES play an insignificant role. This is further demonstrated in Appendix~\ref{app:massagingEES} where the scattering is artificially increased and only a weak effect on transport is seen.

\begin{figure}[ht]
\begin{center}
\includegraphics*[width=\columnwidth]{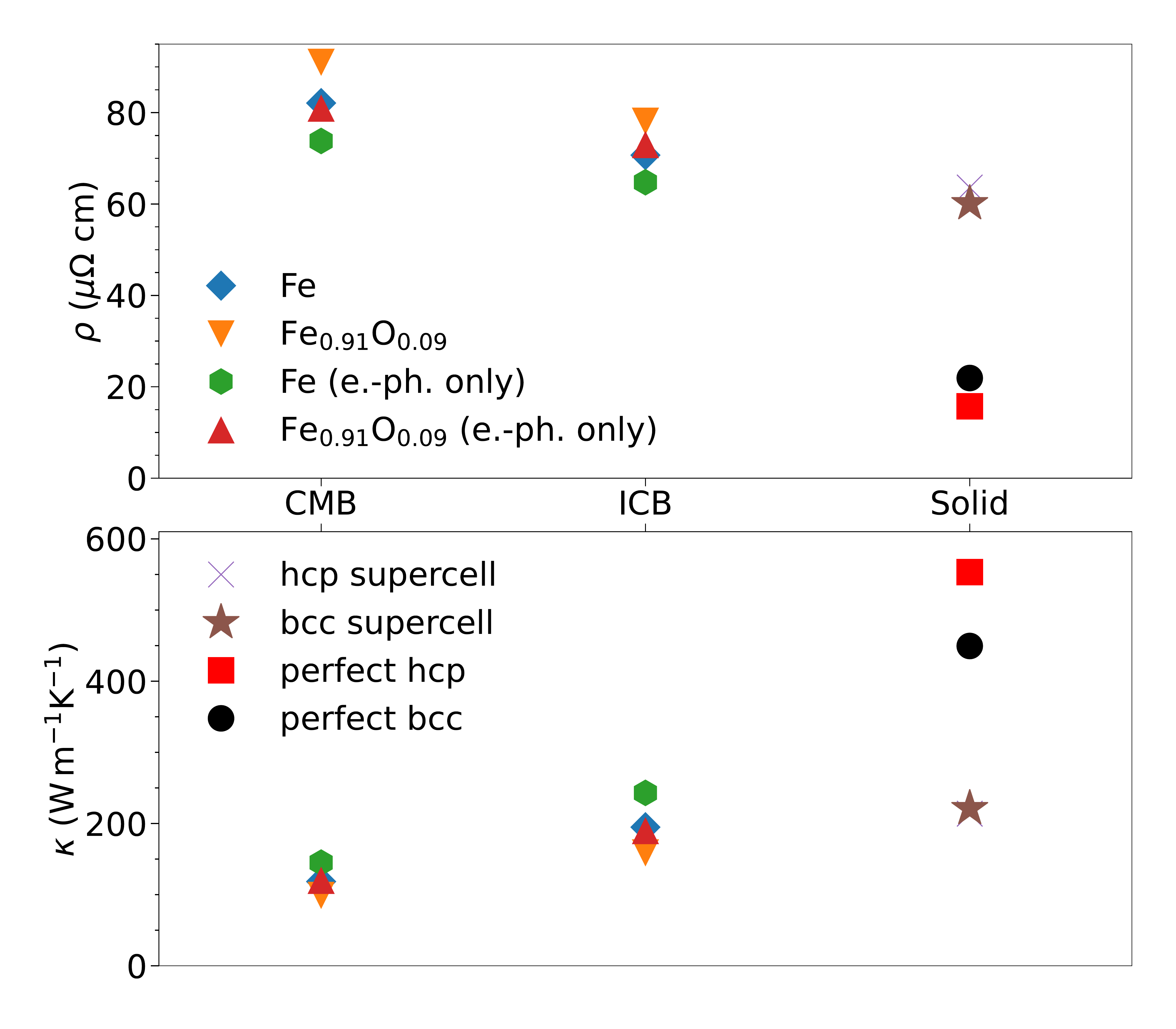}
\caption{Resistivity (top) and thermal conductivity (bottom) for Fe
 and Fe$_{0.91}$O$_{0.09}$   for the ICB and CMB cases calculated using DMFT (including both electron-electron and e.-ph scattering) and DFT (e.-ph. only). Previous calculations (extracted from Ref.~\cite{Pourovskii2020}) for supercell and perfect lattice bcc/hcp are shown for the solid case. } 
\label{fig:resistivity}
\end{center}
\end{figure}

Fig.~\ref{fig:resistivity} shows the calculated values of resistivity (top) and thermal conductivity (bottom) along with data from the literature. Thermal disorder dominates, and perfect crystalline lattices have much higher conductivities. The additional influence of compositional disorder is moderate, and the magnitude of the change is similar to that of including EES on top of the thermal disorder for a given composition.

{\bf Suppression of Lorenz ratio.}
\begin{figure}[ht]
\begin{center}
\includegraphics*[width=0.95\columnwidth]{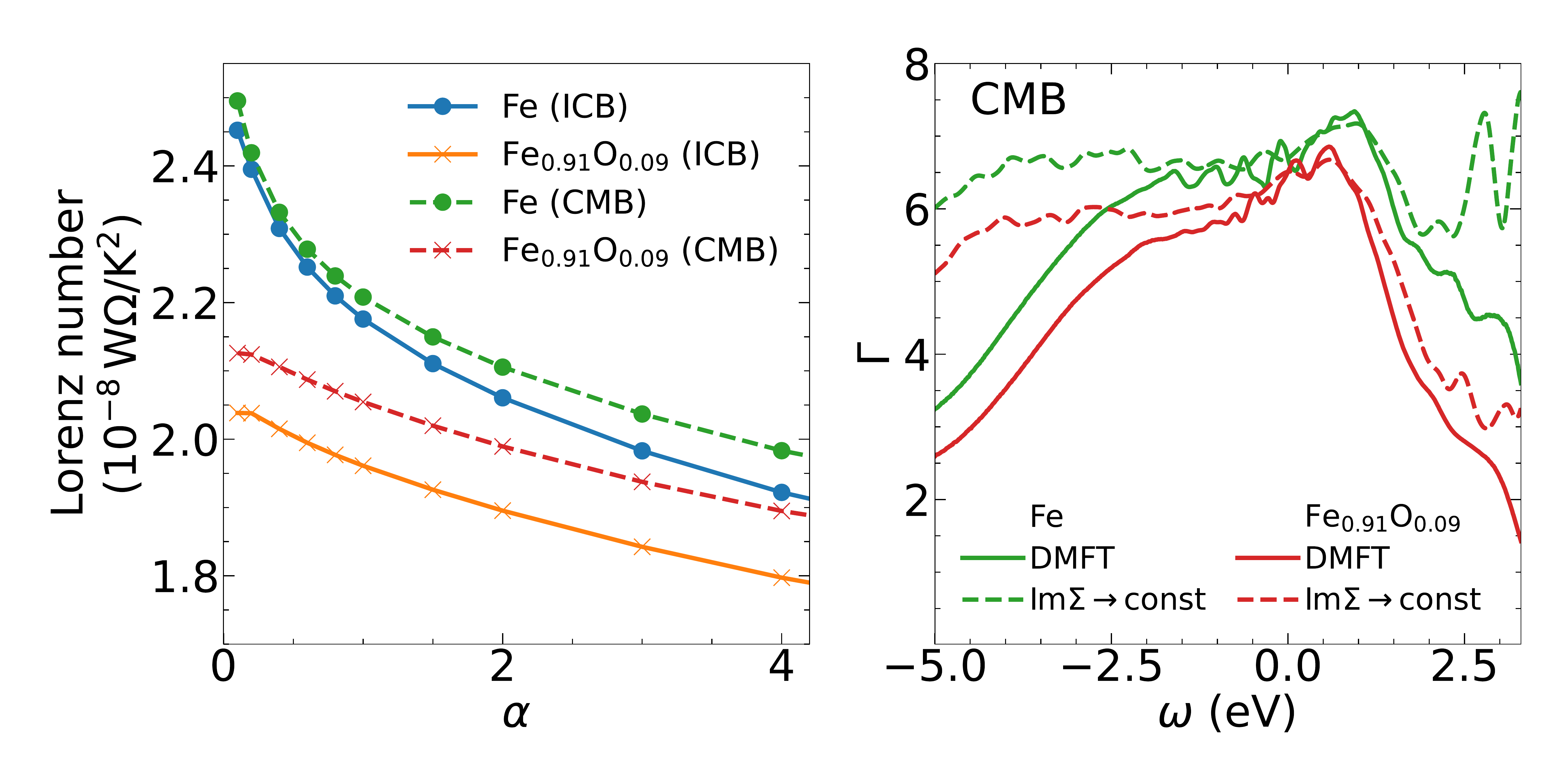}
\caption{(left) Lorenz number for Fe 
 (bullets) and Fe$_{0.91}$O$_{0.09}$
  (crosses) and for the ICB (full line) and CMB cases (dashed line). We use the parameter $\alpha$ to artificially change the magnitude of the EES $\Sigma \rightarrow \Sigma_\alpha= \mathrm{Re} \Sigma +  \alpha i  \mathrm{Im} \Sigma$. One sees that the Lorenz number is reduced by the presence of oxygen. (right) Transport distribution $\Gamma(\omega)$. Solid curves are the obtained form the full DMFT self energy, dashed lines from a constant scattering rate approximation.
}
\label{fig:lorenz}
\end{center}
\end{figure}
Interestingly, EES suppresses the thermal conductivities more than the electrical ones.  Figure \ref{fig:lorenz} shows the evolution of the Lorenz number $L=\kappa/(\sigma T)$ with respect to the strength of EES, which is scaled by the factor $\alpha$, as described in Appendix \ref{app:massagingEES}. For pure Fe, the value of $L$ due to electron-phonon scattering is almost identical to the standard value of 2.44$\cdot 10^{-8}$ (W$\Omega$/K$^2$), whereas it is considerably reduced when EES is taken into account. This reduction occurs because inelastic EES affects $\kappa$ more strongly than $\sigma$. Specifically, $\kappa$ is determined by integrating $\Gamma(\omega) \omega^2 (-df/d\omega)$, where $\Gamma(\omega)$ is the transport distribution function, $f$ is the Fermi function, and $\omega$ is the frequency. On the other hand, $\sigma$ is calculated by integrating $\Gamma(\omega) (-df/d\omega)$, which only involves the derivative of the Fermi function. The conductivity is mostly given by  states  around $\omega=0$, while the dominant contribution to thermal conductivity occurs at finite energies $ 1.5T \lesssim |\omega| \lesssim 4T$.

The right panel of Figure \ref{fig:lorenz} presents the transport distribution $\Gamma(\omega)$ for the CMB case, which are evaluated for the actual EES (full) and compared to a calculation where the energy dependence of scattering is suppressed and the self-energy Im$\Sigma \rightarrow$const is taken (dashed), corresponding to a DFT transport distribution. It is evident that the increase of EES with energy suppresses $\Gamma$ that becomes smaller at larger energies compared to the DFT transport distribution case. Additionally, a comparison between Fe and Fe-O is interesting, which is evident from the DFT transport distributions. The inclusion of oxygen leads to significant suppression of transport distribution at $\omega=2.5$eV, caused by O-2p hybridization with  the 4s iron states. This explains the smaller Lorenz number in the oxygen-rich case.

\section*{Discussion}
\begin{table}[]
\begin{tabular}{l|ll|ll|}
\cline{2-5}
                       & \multicolumn{2}{c|}{{\bf CMB}}    & \multicolumn{2}{c|}{{\bf ICB}}    \\ 
                       & \multicolumn{2}{c|}{$T=$4400~K}   &  \multicolumn{2}{c|}{$T=$6350~K}   \\
                      &  \multicolumn{2}{c|}{$V=$8.64 \AA$^3$/at} &
                          \multicolumn{2}{c|}{$V=$7.16 \AA$^3$/at}   \\ \hline
\multicolumn{1}{|l|}{Case} & \multicolumn{1}{l|}{Fe} & Fe$_{0.91}$O$_{0.09}$ & \multicolumn{1}{l|}{Fe} & Fe$_{0.91}$O$_{0.09}$ \\ \hline
\multicolumn{1}{|l|}{$-\mathrm{Im}\langle\Sigma(\omega=0)\rangle\, $(eV)} & \multicolumn{1}{l|}{0.11} & 0.14 & \multicolumn{1}{l|}{0.09} & 0.11 \\ \hline
\multicolumn{1}{|l|}{$\sigma_{\mathrm{DFT}}$  $~(10^4 \Omega^{-1}\,$cm$^{-1})$} & \multicolumn{1}{l|}{1.35} & 1.24  & \multicolumn{1}{l|}{1.54} & 1.37 \\ 
\multicolumn{1}{|l|}{$\sigma_{\mathrm{DMFT}}$ $~(10^4 \Omega^{-1}$\,cm$^{-1})$} & \multicolumn{1}{l|}{1.22} & 1.10 & \multicolumn{1}{l|}{1.41} & 1.28 \\ 
\multicolumn{1}{|l|}{$1-\sigma_{\mathrm{DMFT}}/\sigma_{\mathrm{DFT}}~ (\%)$} & \multicolumn{1}{l|}{10} & 11  & \multicolumn{1}{l|}{9} & 7 \\ \hline
\multicolumn{1}{|l|}{$\kappa_{\mathrm{DFT}}\,($W$\, $m$^{-1}$K$^{-1}) $} & \multicolumn{1}{l|}{145} & 120  & \multicolumn{1}{l|}{243} & 190 \\ 
\multicolumn{1}{|l|}{$\kappa_{\mathrm{DMFT}}\,($W$\, $m$^{-1}$K$^{-1}) $} & \multicolumn{1}{l|}{119} & 99 & \multicolumn{1}{l|}{193} & 158 \\ 
\multicolumn{1}{|l|}{$1-\kappa_{\mathrm{DMFT}}/\kappa_{\mathrm{DFT}}~ (\%)$} & \multicolumn{1}{l|}{18} & 17  & \multicolumn{1}{l|}{21} & 17 \\  \hline
\multicolumn{1}{|l|}{$L_{\mathrm{DFT}}\,(10^{-8}$ W$\Omega\, $K$^{-2}) $} & \multicolumn{1}{l|}{2.43} & 2.20  & \multicolumn{1}{l|}{2.48} & 2.18 \\ 
\multicolumn{1}{|l|}{$L_{\mathrm{DMFT}}\,(10^{-8}$ W$\Omega\, $K$^{-2}) $} & \multicolumn{1}{l|}{2.22} & 2.05 & \multicolumn{1}{l|}{2.16} & 1.95  \\ \hline
\end{tabular}
\caption{Summary of results for CMB and ICB conditions, both for Fe and Fe$_{0.91}$O$_{0.09}$. 
}
\label{table1}
\end{table}

In summary, our study focused on the impact of oxygen on electronic
transport in liquid iron corresponding to the outer Earth's core
conditions. Because oxygen diminishes the Fe 3$d$ electronic
occupation towards half-filling, it can be expected to strongly
enhance electronic correlations. We indeed find that the EES is
moderately increased.  The numerical values of the conductivities and
the Lorenz ratio are given in Table~\ref{table1}.  The oxygen
substitution at $\sim 10\%$ level, which is a plausible content for
the outer core, suppresses electrical (thermal) conductivities by about
10(17)$\%$ only. In both Fe and Fe-O including electronic correlations diminishes electrical conductivities by less than 10\% and thermal conductivities by about 20$\%$. This reduction is consistent with similar studies~\cite{Pourovskii2020,Zhang2022}, which suggests that it can be used as a rule of thumb  when direct calculations are not feasible.

What are the geophysical implications of electronic correlations? The first observation is that their effect is moderate. The drastic reduction of conductivity, which was predicted on the basis of EES enhancement in ordered Fe-O structures \cite{Zhang2022}, is not observed when thermal disorder effects are simultaneously included.  But it is also clear from our study that neither can one neglect electronic correlations, since the reduction of conductivities due to EES is comparable to the reduction of electron-phonon scattering due to light elements. Importantly, some models that assume a higher heat flow from the core find a thermally only driven geodynamo for $\kappa$ below a limiting value that is of order 100W/mK~\cite{Driscoll2014,Landeau2022}. EES whereas small compared to the thermal disorder, might in the end provide just the necessary additional scattering next to the compositional disorder to power the geodynamo sufficiently. At the very least, whenever one argues the compositional disorder is important, EES must not be neglected either.  Another important finding is the universal suppression of Lorenz number: thermal conductivities are affected by EES more than electrical ones~\cite{Pourovskii2017,Xu2018}. This is seen to be also an effect of compositional disorder in the Fe-O case but EES enhances this further, by supressing the contribution of states away from the Fermi energy. This is important both to properly interpret the high-pressure measurements that mostly probe $\sigma$ and for the geodynamo, because of the distinct influence of the two quantities there.

In future studies, it would be interesting to investigate also alloying with sulfur and silicon~\cite{Wagle2019}. Both elements affect the transport properties strongly at the DFT level because unlike oxygen that alloys interstitially~\cite{Wagle2019}, they alloy substitutionally and therefore more strongly affect the bond disorder with perhaps different implications for electronic correlations. The silicon case was recently investigated~\cite{Zhang2022}, and the results for CMB seem compatible with what we find at comparable concentrations of oxygen, but sulfur that would act also in an oxidizing way could potentially have a bigger effect. Whereas the sulfur concentrations are believed to be negligible in the Earth~\cite{Dreibus1996}, they are expected to be sizable in extraterrestial planets~\cite{Namur2016}. Finaly, both thermal disorder and compositional disorder are also important for Fe oxides that are relevant for the properties of the lower mantle, where, for example, FeO is predicted to be~\cite{ohta2012,leonov2017,leonov2020,WaiGa2023} in  a state where electronic correlations are very strong and the influence of thermal and compositional disorder might lead to large effects there.

\section*{Acknowledgments}
GGB and JM are supported by Slovenian Research Agency (ARRS) under
Grant no. P1-0044 and J1-2458. JM acknowledges discussions with
A. Georges.  DA and MP are supported by the U.K. Natural Environment
Research Council (NERC) under Grant no. NE/T000228/1 and
NE/R000425/1. Computations were performed on the supercomputer Vega at
the Institute of Information Science (IZUM) in Maribor, Slovenia and
in UK on the UK national service Archer2.

\section*{Author contributions}
G.B.B. carried out the DFT + DMFT electronic structure and transport `
calculations. D.A. and M.P. carried out the DFT molecular dynamics and
transport calculations. G.G.B., J.M, L.V.P, M.A, and D.A discussed the
results and wrote the paper.

\section*{Competing Interests}
The authors declare no competing interests.

\appendix
\section{Matsubara self-energies}
\begin{figure}[ht]
\begin{center}
\includegraphics*[width=0.8\columnwidth]{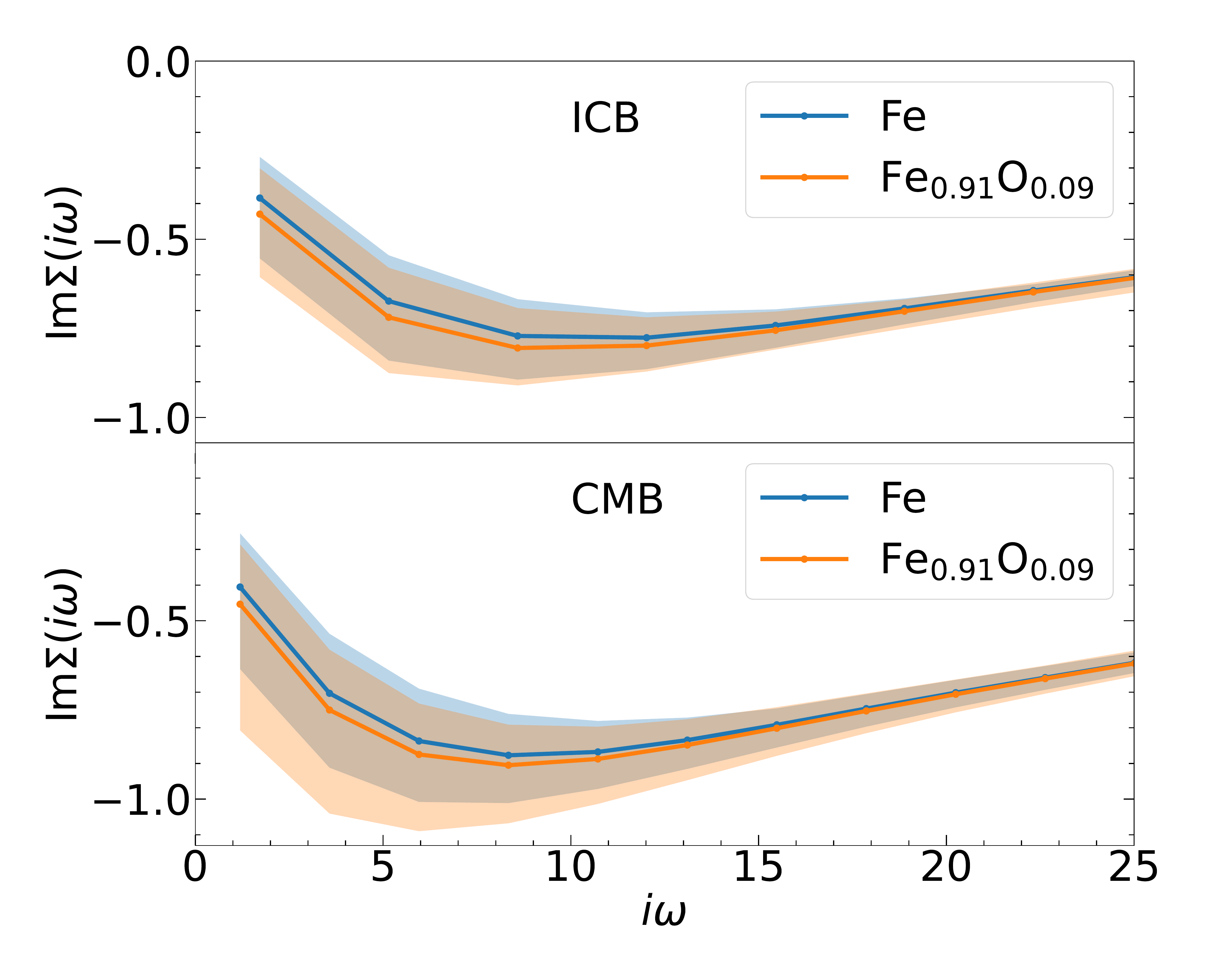}
\caption{Imaginary part of the Matsubara self-energy for the Fe 
 (blue) and Fe$_{0.91}$O$_{0.09}$
  (orange) configurations and ICB (top) and CMB cases (bottom). The lines represent the average over all sites, orbitals, and spins, while the shadow area shows the range of values. For both temperatures, the configuration with oxygen is below the pure iron one.}
\label{fig:self_iw}
\end{center}
\end{figure}

Figure~\ref{fig:self_iw} depicts the imaginary part of the self-energy as a function of Matsubara frequency at ICB (top) and CMB (bottom). The site, spin, and orbital average is performed and the corresponding self-energy is indicated by a full line, whereas the individual self-energies smoothly span the full range, indicated by shading.  One sees the enhancement of EES for the oxygen-rich case in terms of larger magnitude (i.e. more negative values) of $\Sigma(i \omega)$.  

\section{Effects of increased electron-electron scattering}
\label{app:massagingEES}
\begin{figure}[ht]
\begin{center}
  \includegraphics*[width=0.7\columnwidth]{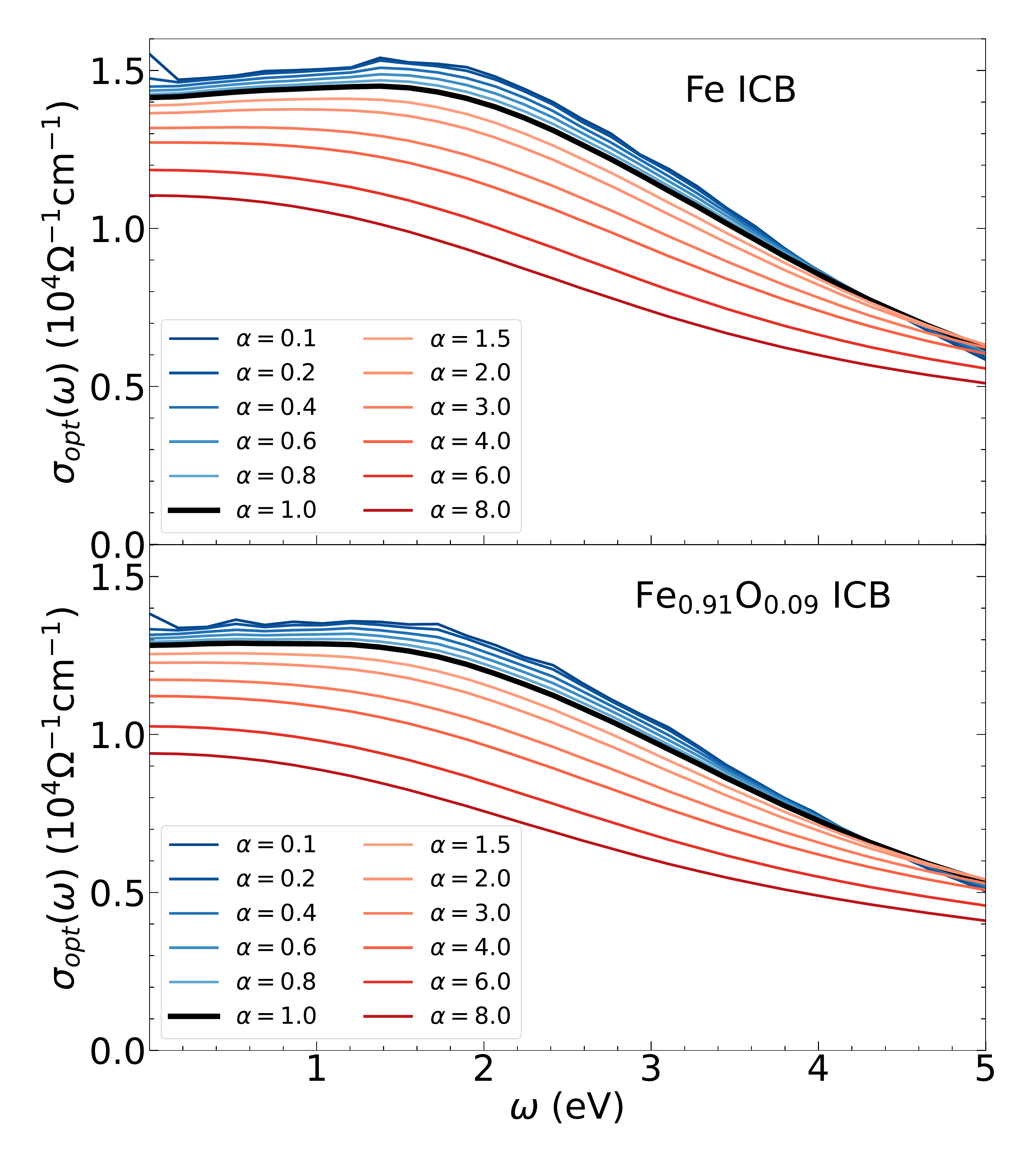}
    \includegraphics*[width=0.7\columnwidth]{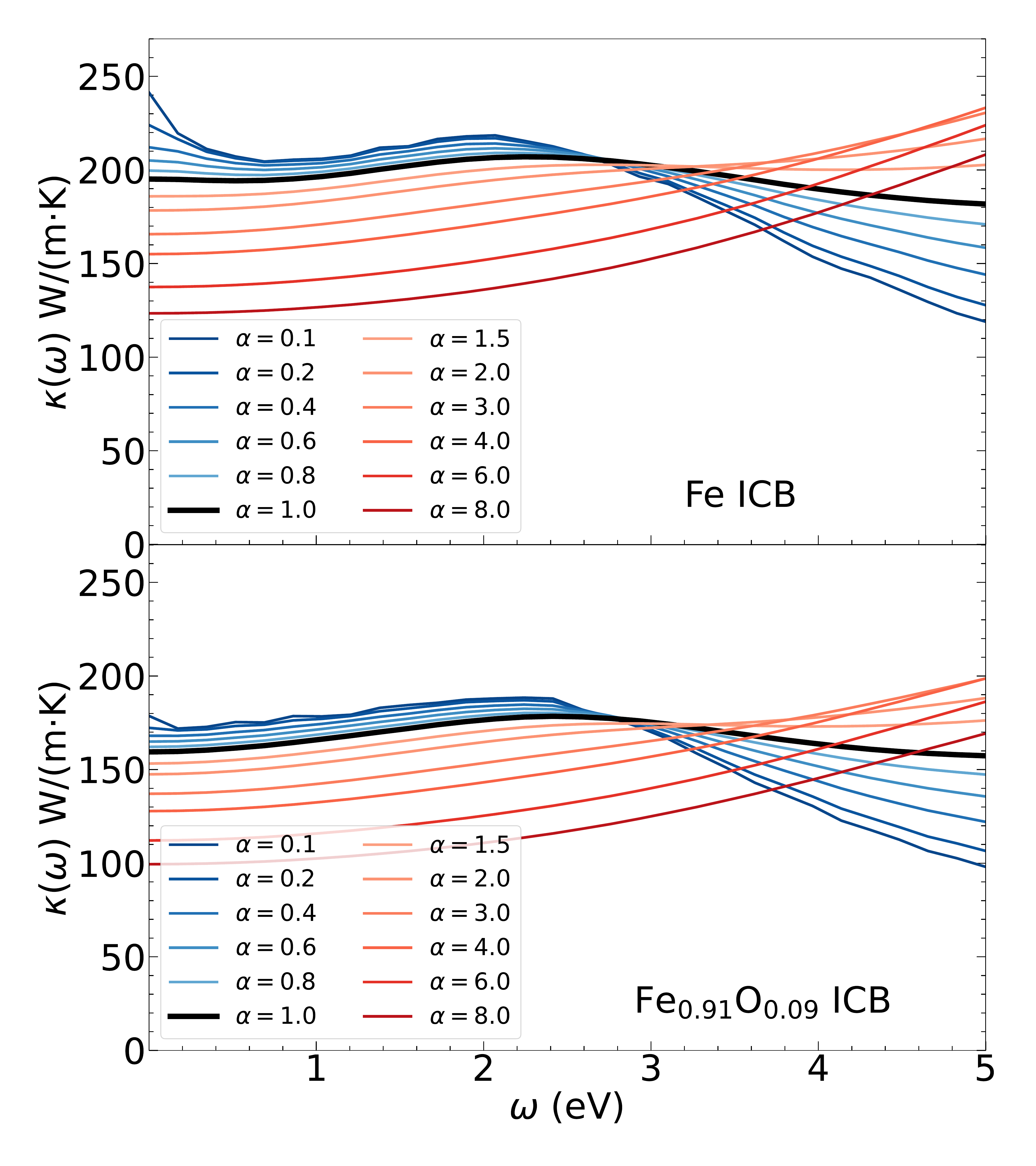}
  
\caption{The optical conductivity and thermal optical conductivity of Fe
 and Fe$_{0.91}$O$_{0.09}$ 
  configuration with a scale factor $\alpha$ applied to the imaginary part of calculated average self-energy. }
\label{fig:alpha_opt}
\end{center}
\end{figure}

In order to investigate how results depend on the strength of EES we performed  trial conductivity calculation where we artificially set $\Sigma \rightarrow \Sigma_\alpha= \mathrm{Re} \Sigma +  \alpha i  \mathrm{Im} \Sigma$, thus increasing the scattering by a scale factor $\alpha$. The calculated optical conductivities and thermal optical conductivities are shown in Fig.~\ref{fig:alpha_opt}. One sees a weak dependence on $\alpha$ around $\alpha=1$.  
As discussed in Ref.~\cite{Pourovskii2020}, the dependence of conductivities on the EES is weaker than what one would expect from  Matthiessen's rule. 
The breakdown of Matthiessen's rule was also documented as a function of the substitutional disorder~\cite{Gomi2016}.

\begin{thebibliography}{51}%
\makeatletter
\providecommand \@ifxundefined [1]{%
 \@ifx{#1\undefined}
}%
\providecommand \@ifnum [1]{%
 \ifnum #1\expandafter \@firstoftwo
 \else \expandafter \@secondoftwo
 \fi
}%
\providecommand \@ifx [1]{%
 \ifx #1\expandafter \@firstoftwo
 \else \expandafter \@secondoftwo
 \fi
}%
\providecommand \natexlab [1]{#1}%
\providecommand \enquote  [1]{``#1''}%
\providecommand \bibnamefont  [1]{#1}%
\providecommand \bibfnamefont [1]{#1}%
\providecommand \citenamefont [1]{#1}%
\providecommand \href@noop [0]{\@secondoftwo}%
\providecommand \href [0]{\begingroup \@sanitize@url \@href}%
\providecommand \@href[1]{\@@startlink{#1}\@@href}%
\providecommand \@@href[1]{\endgroup#1\@@endlink}%
\providecommand \@sanitize@url [0]{\catcode `\\12\catcode `\$12\catcode
  `\&12\catcode `\#12\catcode `\^12\catcode `\_12\catcode `\%12\relax}%
\providecommand \@@startlink[1]{}%
\providecommand \@@endlink[0]{}%
\providecommand \url  [0]{\begingroup\@sanitize@url \@url }%
\providecommand \@url [1]{\endgroup\@href {#1}{\urlprefix }}%
\providecommand \urlprefix  [0]{URL }%
\providecommand \Eprint [0]{\href }%
\providecommand \doibase [0]{https://doi.org/}%
\providecommand \selectlanguage [0]{\@gobble}%
\providecommand \bibinfo  [0]{\@secondoftwo}%
\providecommand \bibfield  [0]{\@secondoftwo}%
\providecommand \translation [1]{[#1]}%
\providecommand \BibitemOpen [0]{}%
\providecommand \bibitemStop [0]{}%
\providecommand \bibitemNoStop [0]{.\EOS\space}%
\providecommand \EOS [0]{\spacefactor3000\relax}%
\providecommand \BibitemShut  [1]{\csname bibitem#1\endcsname}%
\let\auto@bib@innerbib\@empty
\bibitem [{\citenamefont {Ohta}\ \emph {et~al.}(2016)\citenamefont {Ohta},
  \citenamefont {Kuwayama}, \citenamefont {Hirose}, \citenamefont {Shimizu},\
  and\ \citenamefont {Ohishi}}]{Ohta2016}%
  \BibitemOpen
  \bibfield  {author} {\bibinfo {author} {\bibfnamefont {K.}~\bibnamefont
  {Ohta}}, \bibinfo {author} {\bibfnamefont {Y.}~\bibnamefont {Kuwayama}},
  \bibinfo {author} {\bibfnamefont {K.}~\bibnamefont {Hirose}}, \bibinfo
  {author} {\bibfnamefont {K.}~\bibnamefont {Shimizu}},\ and\ \bibinfo {author}
  {\bibfnamefont {Y.}~\bibnamefont {Ohishi}},\ }\bibfield  {title} {\bibinfo
  {title} {Experimental determination of the electrical resistivity of iron at
  earth’s core conditions},\ }\href@noop {} {\bibfield  {journal} {\bibinfo
  {journal} {Nature}\ }\textbf {\bibinfo {volume} {534}},\ \bibinfo {pages}
  {95} (\bibinfo {year} {2016})}\BibitemShut {NoStop}%
\bibitem [{\citenamefont {Kon\^{o}pkov\'a}\ \emph {et~al.}(2016)\citenamefont
  {Kon\^{o}pkov\'a}, \citenamefont {McWilliams}, \citenamefont
  {G\'omez-Pérez},\ and\ \citenamefont {Goncharov}}]{Konopkova2016}%
  \BibitemOpen
  \bibfield  {author} {\bibinfo {author} {\bibfnamefont {Z.}~\bibnamefont
  {Kon\^{o}pkov\'a}}, \bibinfo {author} {\bibfnamefont {R.~S.}\ \bibnamefont
  {McWilliams}}, \bibinfo {author} {\bibfnamefont {N.}~\bibnamefont
  {G\'omez-Pérez}},\ and\ \bibinfo {author} {\bibfnamefont {A.~F.}\
  \bibnamefont {Goncharov}},\ }\bibfield  {title} {\bibinfo {title} {Direct
  measurement of thermal conductivity in solid iron at planetary core
  conditions},\ }\href@noop {} {\bibfield  {journal} {\bibinfo  {journal}
  {Nature (London)}\ }\textbf {\bibinfo {volume} {534}},\ \bibinfo {pages} {99}
  (\bibinfo {year} {2016})}\BibitemShut {NoStop}%
\bibitem [{\citenamefont {Lobanov}\ and\ \citenamefont
  {Geballe}(2022)}]{Lobanov2022}%
  \BibitemOpen
  \bibfield  {author} {\bibinfo {author} {\bibfnamefont {S.~S.}\ \bibnamefont
  {Lobanov}}\ and\ \bibinfo {author} {\bibfnamefont {Z.~M.}\ \bibnamefont
  {Geballe}},\ }\bibfield  {title} {\bibinfo {title} {Non-isotropic contraction
  and expansion of samples in diamond anvil cells: Implications for thermal
  conductivity at the core-mantle boundary},\ }\bibfield  {journal} {\bibinfo
  {journal} {Geophysical Research Letters}\ }\textbf {\bibinfo {volume} {49}},\
  \href {https://doi.org/10.1029/2022gl100379} {10.1029/2022gl100379} (\bibinfo
  {year} {2022})\BibitemShut {NoStop}%
\bibitem [{\citenamefont {Alf\`e}\ and\ \citenamefont
  {Gillan}(1998)}]{Alfe1998}%
  \BibitemOpen
  \bibfield  {author} {\bibinfo {author} {\bibfnamefont {D.}~\bibnamefont
  {Alf\`e}}\ and\ \bibinfo {author} {\bibfnamefont {M.~J.}\ \bibnamefont
  {Gillan}},\ }\bibfield  {title} {\bibinfo {title} {First-principles
  simulations of liquid fe-s under earth's core conditions},\ }\href
  {https://doi.org/10.1103/PhysRevB.58.8248} {\bibfield  {journal} {\bibinfo
  {journal} {Phys. Rev. B}\ }\textbf {\bibinfo {volume} {58}},\ \bibinfo
  {pages} {8248} (\bibinfo {year} {1998})}\BibitemShut {NoStop}%
\bibitem [{\citenamefont {Pozzo}\ \emph {et~al.}(2012)\citenamefont {Pozzo},
  \citenamefont {Davies}, \citenamefont {Gubbins},\ and\ \citenamefont
  {Alf\`e}}]{Pozzo2012}%
  \BibitemOpen
  \bibfield  {author} {\bibinfo {author} {\bibfnamefont {M.}~\bibnamefont
  {Pozzo}}, \bibinfo {author} {\bibfnamefont {C.}~\bibnamefont {Davies}},
  \bibinfo {author} {\bibfnamefont {D.}~\bibnamefont {Gubbins}},\ and\ \bibinfo
  {author} {\bibfnamefont {D.}~\bibnamefont {Alf\`e}},\ }\bibfield  {title}
  {\bibinfo {title} {Thermal and electrical conductivity of iron at earth's
  core conditions},\ }\href {https://doi.org/10.1038/nature11031} {\bibfield
  {journal} {\bibinfo  {journal} {Nature}\ }\textbf {\bibinfo {volume} {485}},\
  \bibinfo {pages} {355} (\bibinfo {year} {2012})}\BibitemShut {NoStop}%
\bibitem [{\citenamefont {de~Koker}\ \emph {et~al.}(2012)\citenamefont
  {de~Koker}, \citenamefont {Steinle-Neumann},\ and\ \citenamefont
  {Vl\u{c}ek}}]{deKoker2012}%
  \BibitemOpen
  \bibfield  {author} {\bibinfo {author} {\bibfnamefont {N.}~\bibnamefont
  {de~Koker}}, \bibinfo {author} {\bibfnamefont {G.}~\bibnamefont
  {Steinle-Neumann}},\ and\ \bibinfo {author} {\bibfnamefont {V.}~\bibnamefont
  {Vl\u{c}ek}},\ }\bibfield  {title} {\bibinfo {title} {Electrical resistivity
  and thermal conductivity of liquid fe alloys at high p and t, and heat flux
  in earth's core},\ }\href@noop {} {\bibfield  {journal} {\bibinfo  {journal}
  {Proc. Natl. Acad. Sci. U. S. A.}\ }\textbf {\bibinfo {volume} {109}},\
  \bibinfo {pages} {4070} (\bibinfo {year} {2012})}\BibitemShut {NoStop}%
\bibitem [{\citenamefont {Stacey}\ and\ \citenamefont
  {Anderson}(2001)}]{Stacey2001}%
  \BibitemOpen
  \bibfield  {author} {\bibinfo {author} {\bibfnamefont {F.~D.}\ \bibnamefont
  {Stacey}}\ and\ \bibinfo {author} {\bibfnamefont {O.~L.}\ \bibnamefont
  {Anderson}},\ }\bibfield  {title} {\bibinfo {title} {Electrical and thermal
  conductivities of fe{\textendash}ni{\textendash}si alloy under core
  conditions},\ }\href {https://doi.org/10.1016/s0031-9201(01)00186-8}
  {\bibfield  {journal} {\bibinfo  {journal} {Physics of the Earth and
  Planetary Interiors}\ }\textbf {\bibinfo {volume} {124}},\ \bibinfo {pages}
  {153} (\bibinfo {year} {2001})}\BibitemShut {NoStop}%
\bibitem [{\citenamefont {Stacey}\ and\ \citenamefont
  {Loper}(2007)}]{Stacey2007}%
  \BibitemOpen
  \bibfield  {author} {\bibinfo {author} {\bibfnamefont {F.}~\bibnamefont
  {Stacey}}\ and\ \bibinfo {author} {\bibfnamefont {D.}~\bibnamefont {Loper}},\
  }\bibfield  {title} {\bibinfo {title} {A revised estimate of the conductivity
  of iron alloy at high pressure and implications for the core energy
  balance},\ }\href@noop {} {\bibfield  {journal} {\bibinfo  {journal} {Physics
  of the Earth and Planetary Interiors}\ }\textbf {\bibinfo {volume} {161}},\
  \bibinfo {pages} {13} (\bibinfo {year} {2007})}\BibitemShut {NoStop}%
\bibitem [{\citenamefont {Gunnarsson}\ \emph {et~al.}(2003)\citenamefont
  {Gunnarsson}, \citenamefont {Calandra},\ and\ \citenamefont
  {Han}}]{Gunnarsson2003}%
  \BibitemOpen
  \bibfield  {author} {\bibinfo {author} {\bibfnamefont {O.}~\bibnamefont
  {Gunnarsson}}, \bibinfo {author} {\bibfnamefont {M.}~\bibnamefont
  {Calandra}},\ and\ \bibinfo {author} {\bibfnamefont {J.~E.}\ \bibnamefont
  {Han}},\ }\bibfield  {title} {\bibinfo {title} {Colloquium: Saturation of
  electrical resistivity},\ }\href {https://doi.org/10.1103/RevModPhys.75.1085}
  {\bibfield  {journal} {\bibinfo  {journal} {Rev. Mod. Phys.}\ }\textbf
  {\bibinfo {volume} {75}},\ \bibinfo {pages} {1085} (\bibinfo {year}
  {2003})}\BibitemShut {NoStop}%
\bibitem [{\citenamefont {Gomi}\ \emph {et~al.}(2013)\citenamefont {Gomi},
  \citenamefont {Ohta}, \citenamefont {Hirose}, \citenamefont {Labrosse},
  \citenamefont {Caracas}, \citenamefont {Verstraete},\ and\ \citenamefont
  {Hernlund}}]{Gomi2013}%
  \BibitemOpen
  \bibfield  {author} {\bibinfo {author} {\bibfnamefont {H.}~\bibnamefont
  {Gomi}}, \bibinfo {author} {\bibfnamefont {K.}~\bibnamefont {Ohta}}, \bibinfo
  {author} {\bibfnamefont {K.}~\bibnamefont {Hirose}}, \bibinfo {author}
  {\bibfnamefont {S.}~\bibnamefont {Labrosse}}, \bibinfo {author}
  {\bibfnamefont {R.}~\bibnamefont {Caracas}}, \bibinfo {author} {\bibfnamefont
  {M.~J.}\ \bibnamefont {Verstraete}},\ and\ \bibinfo {author} {\bibfnamefont
  {J.~W.}\ \bibnamefont {Hernlund}},\ }\bibfield  {title} {\bibinfo {title}
  {The high conductivity of iron and thermal evolution of the earth's core},\
  }\href {https://doi.org/10.1016/j.pepi.2013.07.010} {\bibfield  {journal}
  {\bibinfo  {journal} {Physics of the Earth and Planetary Interiors}\ }\textbf
  {\bibinfo {volume} {224}},\ \bibinfo {pages} {88} (\bibinfo {year}
  {2013})}\BibitemShut {NoStop}%
\bibitem [{\citenamefont {Gomi}\ \emph {et~al.}(2016)\citenamefont {Gomi},
  \citenamefont {Hirose}, \citenamefont {Akai},\ and\ \citenamefont
  {Fei}}]{Gomi2016}%
  \BibitemOpen
  \bibfield  {author} {\bibinfo {author} {\bibfnamefont {H.}~\bibnamefont
  {Gomi}}, \bibinfo {author} {\bibfnamefont {K.}~\bibnamefont {Hirose}},
  \bibinfo {author} {\bibfnamefont {H.}~\bibnamefont {Akai}},\ and\ \bibinfo
  {author} {\bibfnamefont {Y.}~\bibnamefont {Fei}},\ }\bibfield  {title}
  {\bibinfo {title} {Electrical resistivity of substitutionally disordered hcp
  fe{\textendash}si and fe{\textendash}ni alloys: Chemically-induced
  resistivity saturation in the earth{\textquotesingle}s core},\ }\href
  {https://doi.org/10.1016/j.epsl.2016.07.011} {\bibfield  {journal} {\bibinfo
  {journal} {Earth and Planetary Science Letters}\ }\textbf {\bibinfo {volume}
  {451}},\ \bibinfo {pages} {51} (\bibinfo {year} {2016})}\BibitemShut
  {NoStop}%
\bibitem [{\citenamefont {Tarduno}\ \emph {et~al.}(2010)\citenamefont
  {Tarduno}, \citenamefont {Cottrell}, \citenamefont {Watkeys}, \citenamefont
  {Hofmann}, \citenamefont {Doubrovine}, \citenamefont {Mamajek}, \citenamefont
  {Liu}, \citenamefont {Sibeck}, \citenamefont {Neukirch},\ and\ \citenamefont
  {Usui}}]{Tarduno2010}%
  \BibitemOpen
  \bibfield  {author} {\bibinfo {author} {\bibfnamefont {J.~A.}\ \bibnamefont
  {Tarduno}}, \bibinfo {author} {\bibfnamefont {R.~D.}\ \bibnamefont
  {Cottrell}}, \bibinfo {author} {\bibfnamefont {M.~K.}\ \bibnamefont
  {Watkeys}}, \bibinfo {author} {\bibfnamefont {A.}~\bibnamefont {Hofmann}},
  \bibinfo {author} {\bibfnamefont {P.~V.}\ \bibnamefont {Doubrovine}},
  \bibinfo {author} {\bibfnamefont {E.~E.}\ \bibnamefont {Mamajek}}, \bibinfo
  {author} {\bibfnamefont {D.}~\bibnamefont {Liu}}, \bibinfo {author}
  {\bibfnamefont {D.~G.}\ \bibnamefont {Sibeck}}, \bibinfo {author}
  {\bibfnamefont {L.~P.}\ \bibnamefont {Neukirch}},\ and\ \bibinfo {author}
  {\bibfnamefont {Y.}~\bibnamefont {Usui}},\ }\bibfield  {title} {\bibinfo
  {title} {Geodynamo, solar wind, and magnetopause 3.4 to 3.45 billion years
  ago},\ }\href {https://doi.org/10.1126/science.1183445} {\bibfield  {journal}
  {\bibinfo  {journal} {Science}\ }\textbf {\bibinfo {volume} {327}},\ \bibinfo
  {pages} {1238} (\bibinfo {year} {2010})}\BibitemShut {NoStop}%
\bibitem [{\citenamefont {Tarduno}\ \emph {et~al.}(2015)\citenamefont
  {Tarduno}, \citenamefont {Cottrell}, \citenamefont {Davis}, \citenamefont
  {Nimmo},\ and\ \citenamefont {Bono}}]{Tarduno2015}%
  \BibitemOpen
  \bibfield  {author} {\bibinfo {author} {\bibfnamefont {J.~A.}\ \bibnamefont
  {Tarduno}}, \bibinfo {author} {\bibfnamefont {R.~D.}\ \bibnamefont
  {Cottrell}}, \bibinfo {author} {\bibfnamefont {W.~J.}\ \bibnamefont {Davis}},
  \bibinfo {author} {\bibfnamefont {F.}~\bibnamefont {Nimmo}},\ and\ \bibinfo
  {author} {\bibfnamefont {R.~K.}\ \bibnamefont {Bono}},\ }\bibfield  {title}
  {\bibinfo {title} {A hadean to paleoarchean geodynamo recorded by single
  zircon crystals},\ }\href@noop {} {\bibfield  {journal} {\bibinfo  {journal}
  {Science}\ }\textbf {\bibinfo {volume} {349}},\ \bibinfo {pages} {521}
  (\bibinfo {year} {2015})}\BibitemShut {NoStop}%
\bibitem [{\citenamefont {Olson}(2013)}]{Olson2013}%
  \BibitemOpen
  \bibfield  {author} {\bibinfo {author} {\bibfnamefont {P.}~\bibnamefont
  {Olson}},\ }\bibfield  {title} {\bibinfo {title} {The new core paradox},\
  }\href {https://doi.org/10.1126/science.1243477} {\bibfield  {journal}
  {\bibinfo  {journal} {Science}\ }\textbf {\bibinfo {volume} {342}},\ \bibinfo
  {pages} {431} (\bibinfo {year} {2013})}\BibitemShut {NoStop}%
\bibitem [{\citenamefont {Pourovskii}\ \emph {et~al.}(2013)\citenamefont
  {Pourovskii}, \citenamefont {Miyake}, \citenamefont {Simak}, \citenamefont
  {Ruban}, \citenamefont {Dubrovinsky},\ and\ \citenamefont
  {Abrikosov}}]{pourovskii2013}%
  \BibitemOpen
  \bibfield  {author} {\bibinfo {author} {\bibfnamefont {L.}~\bibnamefont
  {Pourovskii}}, \bibinfo {author} {\bibfnamefont {T.}~\bibnamefont {Miyake}},
  \bibinfo {author} {\bibfnamefont {S.}~\bibnamefont {Simak}}, \bibinfo
  {author} {\bibfnamefont {A.~V.}\ \bibnamefont {Ruban}}, \bibinfo {author}
  {\bibfnamefont {L.}~\bibnamefont {Dubrovinsky}},\ and\ \bibinfo {author}
  {\bibfnamefont {I.}~\bibnamefont {Abrikosov}},\ }\bibfield  {title} {\bibinfo
  {title} {Electronic properties and magnetism of iron at the earth's inner
  core conditions},\ }\href@noop {} {\bibfield  {journal} {\bibinfo  {journal}
  {Physical Review B}\ }\textbf {\bibinfo {volume} {87}},\ \bibinfo {pages}
  {115130} (\bibinfo {year} {2013})}\BibitemShut {NoStop}%
\bibitem [{\citenamefont {Pourovskii}(2019)}]{Pourovskii_2019}%
  \BibitemOpen
  \bibfield  {author} {\bibinfo {author} {\bibfnamefont {L.~V.}\ \bibnamefont
  {Pourovskii}},\ }\bibfield  {title} {\bibinfo {title} {Electronic
  correlations in dense iron: from moderate pressure to earth’s core
  conditions},\ }\href {https://doi.org/10.1088/1361-648X/ab274f} {\bibfield
  {journal} {\bibinfo  {journal} {Journal of Physics: Condensed Matter}\
  }\textbf {\bibinfo {volume} {31}},\ \bibinfo {pages} {373001} (\bibinfo
  {year} {2019})}\BibitemShut {NoStop}%
\bibitem [{\citenamefont {Deng}\ \emph {et~al.}(2013)\citenamefont {Deng},
  \citenamefont {Mravlje}, \citenamefont {\ifmmode~\check{Z}\else
  \v{Z}\fi{}itko}, \citenamefont {Ferrero}, \citenamefont {Kotliar},\ and\
  \citenamefont {Georges}}]{Deng2013}%
  \BibitemOpen
  \bibfield  {author} {\bibinfo {author} {\bibfnamefont {X.}~\bibnamefont
  {Deng}}, \bibinfo {author} {\bibfnamefont {J.}~\bibnamefont {Mravlje}},
  \bibinfo {author} {\bibfnamefont {R.}~\bibnamefont {\ifmmode~\check{Z}\else
  \v{Z}\fi{}itko}}, \bibinfo {author} {\bibfnamefont {M.}~\bibnamefont
  {Ferrero}}, \bibinfo {author} {\bibfnamefont {G.}~\bibnamefont {Kotliar}},\
  and\ \bibinfo {author} {\bibfnamefont {A.}~\bibnamefont {Georges}},\
  }\bibfield  {title} {\bibinfo {title} {How bad metals turn good:
  Spectroscopic signatures of resilient quasiparticles},\ }\href
  {https://doi.org/10.1103/PhysRevLett.110.086401} {\bibfield  {journal}
  {\bibinfo  {journal} {Phys. Rev. Lett.}\ }\textbf {\bibinfo {volume} {110}},\
  \bibinfo {pages} {086401} (\bibinfo {year} {2013})}\BibitemShut {NoStop}%
\bibitem [{\citenamefont {Zhang}\ \emph {et~al.}(2015)\citenamefont {Zhang},
  \citenamefont {Cohen},\ and\ \citenamefont {Haule}}]{Zhang2015}%
  \BibitemOpen
  \bibfield  {author} {\bibinfo {author} {\bibfnamefont {P.}~\bibnamefont
  {Zhang}}, \bibinfo {author} {\bibfnamefont {R.}~\bibnamefont {Cohen}},\ and\
  \bibinfo {author} {\bibfnamefont {K.}~\bibnamefont {Haule}},\ }\bibfield
  {title} {\bibinfo {title} {Effects of electron correlations on transport
  properties of iron at earth's core conditions},\ }\href@noop {} {\bibfield
  {journal} {\bibinfo  {journal} {Nature (London)}\ }\textbf {\bibinfo {volume}
  {517}},\ \bibinfo {pages} {605} (\bibinfo {year} {2015})}\BibitemShut
  {NoStop}%
\bibitem [{\citenamefont {Pourovskii}\ \emph {et~al.}(2017)\citenamefont
  {Pourovskii}, \citenamefont {Mravlje}, \citenamefont {Georges}, \citenamefont
  {Simak},\ and\ \citenamefont {Abrikosov}}]{Pourovskii2017}%
  \BibitemOpen
  \bibfield  {author} {\bibinfo {author} {\bibfnamefont {L.~V.}\ \bibnamefont
  {Pourovskii}}, \bibinfo {author} {\bibfnamefont {J.}~\bibnamefont {Mravlje}},
  \bibinfo {author} {\bibfnamefont {A.}~\bibnamefont {Georges}}, \bibinfo
  {author} {\bibfnamefont {S.~I.}\ \bibnamefont {Simak}},\ and\ \bibinfo
  {author} {\bibfnamefont {I.~A.}\ \bibnamefont {Abrikosov}},\ }\bibfield
  {title} {\bibinfo {title} {Electron–electron scattering and thermal
  conductivity of $\epsilon$-iron at earth’s core conditions},\ }\href
  {https://doi.org/10.1088/1367-2630/aa76c9} {\bibfield  {journal} {\bibinfo
  {journal} {New Journal of Physics}\ }\textbf {\bibinfo {volume} {19}},\
  \bibinfo {pages} {073022} (\bibinfo {year} {2017})}\BibitemShut {NoStop}%
\bibitem [{\citenamefont {Xu}\ \emph {et~al.}(2018)\citenamefont {Xu},
  \citenamefont {Zhang}, \citenamefont {Haule}, \citenamefont {Minar},
  \citenamefont {Wimmer}, \citenamefont {Ebert},\ and\ \citenamefont
  {Cohen}}]{Xu2018}%
  \BibitemOpen
  \bibfield  {author} {\bibinfo {author} {\bibfnamefont {J.}~\bibnamefont
  {Xu}}, \bibinfo {author} {\bibfnamefont {P.}~\bibnamefont {Zhang}}, \bibinfo
  {author} {\bibfnamefont {K.}~\bibnamefont {Haule}}, \bibinfo {author}
  {\bibfnamefont {J.}~\bibnamefont {Minar}}, \bibinfo {author} {\bibfnamefont
  {S.}~\bibnamefont {Wimmer}}, \bibinfo {author} {\bibfnamefont
  {H.}~\bibnamefont {Ebert}},\ and\ \bibinfo {author} {\bibfnamefont {R.~E.}\
  \bibnamefont {Cohen}},\ }\bibfield  {title} {\bibinfo {title} {Thermal
  conductivity and electrical resistivity of solid iron at earth's core
  conditions from first principles},\ }\href
  {https://doi.org/10.1103/PhysRevLett.121.096601} {\bibfield  {journal}
  {\bibinfo  {journal} {Phys. Rev. Lett.}\ }\textbf {\bibinfo {volume} {121}},\
  \bibinfo {pages} {096601} (\bibinfo {year} {2018})}\BibitemShut {NoStop}%
\bibitem [{\citenamefont {Pourovskii}\ \emph {et~al.}(2020)\citenamefont
  {Pourovskii}, \citenamefont {Mravlje}, \citenamefont {Pozzo},\ and\
  \citenamefont {Alf\`e}}]{Pourovskii2020}%
  \BibitemOpen
  \bibfield  {author} {\bibinfo {author} {\bibfnamefont {L.~V.}\ \bibnamefont
  {Pourovskii}}, \bibinfo {author} {\bibfnamefont {J.}~\bibnamefont {Mravlje}},
  \bibinfo {author} {\bibfnamefont {M.}~\bibnamefont {Pozzo}},\ and\ \bibinfo
  {author} {\bibfnamefont {D.}~\bibnamefont {Alf\`e}},\ }\bibfield  {title}
  {\bibinfo {title} {Electronic correlations and transport in iron at earth's
  core conditions},\ }\href {https://doi.org/10.1038/s41467-020-18003-9}
  {\bibfield  {journal} {\bibinfo  {journal} {Nature Communications}\ }\textbf
  {\bibinfo {volume} {11}},\ \bibinfo {pages} {4105} (\bibinfo {year}
  {2020})}\BibitemShut {NoStop}%
\bibitem [{\citenamefont {Zhang}\ \emph {et~al.}(2022)\citenamefont {Zhang},
  \citenamefont {Luo}, \citenamefont {Hou}, \citenamefont {Driscoll},
  \citenamefont {Salke}, \citenamefont {Min{\'a}r}, \citenamefont {Prakapenka},
  \citenamefont {Greenberg}, \citenamefont {Hemley}, \citenamefont {Cohen},\
  and\ \citenamefont {Lin}}]{Zhang2022}%
  \BibitemOpen
  \bibfield  {author} {\bibinfo {author} {\bibfnamefont {Y.}~\bibnamefont
  {Zhang}}, \bibinfo {author} {\bibfnamefont {K.}~\bibnamefont {Luo}}, \bibinfo
  {author} {\bibfnamefont {M.}~\bibnamefont {Hou}}, \bibinfo {author}
  {\bibfnamefont {P.}~\bibnamefont {Driscoll}}, \bibinfo {author}
  {\bibfnamefont {N.~P.}\ \bibnamefont {Salke}}, \bibinfo {author}
  {\bibfnamefont {J.}~\bibnamefont {Min{\'a}r}}, \bibinfo {author}
  {\bibfnamefont {V.~B.}\ \bibnamefont {Prakapenka}}, \bibinfo {author}
  {\bibfnamefont {E.}~\bibnamefont {Greenberg}}, \bibinfo {author}
  {\bibfnamefont {R.~J.}\ \bibnamefont {Hemley}}, \bibinfo {author}
  {\bibfnamefont {R.~E.}\ \bibnamefont {Cohen}},\ and\ \bibinfo {author}
  {\bibfnamefont {J.-F.}\ \bibnamefont {Lin}},\ }\bibfield  {title} {\bibinfo
  {title} {Thermal conductivity of fe-si alloys and thermal stratification in
  earth’s core},\ }\href {https://doi.org/10.1073/pnas.2119001119} {\bibfield
   {journal} {\bibinfo  {journal} {Proceedings of the National Academy of
  Sciences}\ }\textbf {\bibinfo {volume} {119}},\ \bibinfo {pages}
  {e2119001119} (\bibinfo {year} {2022})},\ \Eprint
  {https://arxiv.org/abs/https://www.pnas.org/doi/pdf/10.1073/pnas.2119001119}
  {https://www.pnas.org/doi/pdf/10.1073/pnas.2119001119} \BibitemShut {NoStop}%
\bibitem [{\citenamefont {Badro}\ \emph {et~al.}(2015)\citenamefont {Badro},
  \citenamefont {Brodholt}, \citenamefont {Piet}, \citenamefont {Siebert},\
  and\ \citenamefont {Ryerson}}]{Badro2015}%
  \BibitemOpen
  \bibfield  {author} {\bibinfo {author} {\bibfnamefont {J.}~\bibnamefont
  {Badro}}, \bibinfo {author} {\bibfnamefont {J.~P.}\ \bibnamefont {Brodholt}},
  \bibinfo {author} {\bibfnamefont {H.}~\bibnamefont {Piet}}, \bibinfo {author}
  {\bibfnamefont {J.}~\bibnamefont {Siebert}},\ and\ \bibinfo {author}
  {\bibfnamefont {F.~J.}\ \bibnamefont {Ryerson}},\ }\bibfield  {title}
  {\bibinfo {title} {Core formation and core composition from coupled
  geochemical and geophysical constraints},\ }\href
  {https://doi.org/10.1073/pnas.1505672112} {\bibfield  {journal} {\bibinfo
  {journal} {Proceedings of the National Academy of Sciences}\ }\textbf
  {\bibinfo {volume} {112}},\ \bibinfo {pages} {12310} (\bibinfo {year}
  {2015})}\BibitemShut {NoStop}%
\bibitem [{\citenamefont {Wagle}\ \emph {et~al.}(2019)\citenamefont {Wagle},
  \citenamefont {Steinle-Neumann},\ and\ \citenamefont {de~Koker}}]{Wagle2019}%
  \BibitemOpen
  \bibfield  {author} {\bibinfo {author} {\bibfnamefont {F.}~\bibnamefont
  {Wagle}}, \bibinfo {author} {\bibfnamefont {G.}~\bibnamefont
  {Steinle-Neumann}},\ and\ \bibinfo {author} {\bibfnamefont {N.}~\bibnamefont
  {de~Koker}},\ }\bibfield  {title} {\bibinfo {title} {Resistivity saturation
  in liquid iron{\textendash}light-element alloys at conditions of planetary
  cores from first principles computations},\ }\href
  {https://doi.org/10.1016/j.crte.2018.05.002} {\bibfield  {journal} {\bibinfo
  {journal} {Comptes Rendus Geoscience}\ }\textbf {\bibinfo {volume} {351}},\
  \bibinfo {pages} {154} (\bibinfo {year} {2019})}\BibitemShut {NoStop}%
\bibitem [{\citenamefont {Li}\ \emph {et~al.}(2021)\citenamefont {Li},
  \citenamefont {Li}, \citenamefont {He}, \citenamefont {Wang},\ and\
  \citenamefont {Zhang}}]{Li2021}%
  \BibitemOpen
  \bibfield  {author} {\bibinfo {author} {\bibfnamefont {W.-J.}\ \bibnamefont
  {Li}}, \bibinfo {author} {\bibfnamefont {Z.}~\bibnamefont {Li}}, \bibinfo
  {author} {\bibfnamefont {X.-T.}\ \bibnamefont {He}}, \bibinfo {author}
  {\bibfnamefont {C.}~\bibnamefont {Wang}},\ and\ \bibinfo {author}
  {\bibfnamefont {P.}~\bibnamefont {Zhang}},\ }\bibfield  {title} {\bibinfo
  {title} {Constraints on the thermal evolution of earth{\textquotesingle}s
  core from ab initio calculated transport properties of {FeNi} liquids},\
  }\href {https://doi.org/10.1016/j.epsl.2021.116852} {\bibfield  {journal}
  {\bibinfo  {journal} {Earth and Planetary Science Letters}\ }\textbf
  {\bibinfo {volume} {562}},\ \bibinfo {pages} {116852} (\bibinfo {year}
  {2021})}\BibitemShut {NoStop}%
\bibitem [{\citenamefont {Georges}\ \emph {et~al.}(2013)\citenamefont
  {Georges}, \citenamefont {de{\textquotesingle} Medici},\ and\ \citenamefont
  {Mravlje}}]{Georges2013}%
  \BibitemOpen
  \bibfield  {author} {\bibinfo {author} {\bibfnamefont {A.}~\bibnamefont
  {Georges}}, \bibinfo {author} {\bibfnamefont {L.}~\bibnamefont
  {de{\textquotesingle} Medici}},\ and\ \bibinfo {author} {\bibfnamefont
  {J.}~\bibnamefont {Mravlje}},\ }\bibfield  {title} {\bibinfo {title} {Strong
  correlations from hund's coupling},\ }\href
  {https://doi.org/10.1146/annurev-conmatphys-020911-125045} {\bibfield
  {journal} {\bibinfo  {journal} {Annual Review of Condensed Matter Physics}\
  }\textbf {\bibinfo {volume} {4}},\ \bibinfo {pages} {137} (\bibinfo {year}
  {2013})}\BibitemShut {NoStop}%
\bibitem [{\citenamefont {Jang}\ \emph {et~al.}(2021)\citenamefont {Jang},
  \citenamefont {He}, \citenamefont {Shim}, \citenamefont {Mao},\ and\
  \citenamefont {Kim}}]{Jang2021}%
  \BibitemOpen
  \bibfield  {author} {\bibinfo {author} {\bibfnamefont {B.~G.}\ \bibnamefont
  {Jang}}, \bibinfo {author} {\bibfnamefont {Y.}~\bibnamefont {He}}, \bibinfo
  {author} {\bibfnamefont {J.~H.}\ \bibnamefont {Shim}}, \bibinfo {author}
  {\bibfnamefont {H.-k.}\ \bibnamefont {Mao}},\ and\ \bibinfo {author}
  {\bibfnamefont {D.~Y.}\ \bibnamefont {Kim}},\ }\href
  {https://doi.org/10.48550/ARXIV.2111.11033} {\bibinfo {title} {Oxygen-driven
  enhancement of electron correlation in hexagonal iron at earth's inner core
  conditions}} (\bibinfo {year} {2021}),\ \bibinfo {note}
  {arXiv:2111.11033}\BibitemShut {NoStop}%
\bibitem [{\citenamefont {Georges}\ \emph {et~al.}(1996)\citenamefont
  {Georges}, \citenamefont {Kotliar}, \citenamefont {Krauth},\ and\
  \citenamefont {Rozenberg}}]{georges_dmft_1996}%
  \BibitemOpen
  \bibfield  {author} {\bibinfo {author} {\bibfnamefont {A.}~\bibnamefont
  {Georges}}, \bibinfo {author} {\bibfnamefont {G.}~\bibnamefont {Kotliar}},
  \bibinfo {author} {\bibfnamefont {W.}~\bibnamefont {Krauth}},\ and\ \bibinfo
  {author} {\bibfnamefont {M.~J.}\ \bibnamefont {Rozenberg}},\ }\bibfield
  {title} {\bibinfo {title} {Dynamical mean-field theory of strongly correlated
  fermion systems and the limit of infinite dimensions},\ }\href@noop {}
  {\bibfield  {journal} {\bibinfo  {journal} {Reviews of Modern Physics}\
  }\textbf {\bibinfo {volume} {68}},\ \bibinfo {pages} {13} (\bibinfo {year}
  {1996})}\BibitemShut {NoStop}%
\bibitem [{\citenamefont {Hausoel}\ \emph {et~al.}(2017)\citenamefont
  {Hausoel}, \citenamefont {Karolak}, \citenamefont
  {{\c{S}}a{\c{s}}$\upiota$o{\u{g}}lu}, \citenamefont {Lichtenstein},
  \citenamefont {Held}, \citenamefont {Katanin}, \citenamefont {Toschi},\ and\
  \citenamefont {Sangiovanni}}]{Hausoel2017}%
  \BibitemOpen
  \bibfield  {author} {\bibinfo {author} {\bibfnamefont {A.}~\bibnamefont
  {Hausoel}}, \bibinfo {author} {\bibfnamefont {M.}~\bibnamefont {Karolak}},
  \bibinfo {author} {\bibfnamefont {E.}~\bibnamefont
  {{\c{S}}a{\c{s}}$\upiota$o{\u{g}}lu}}, \bibinfo {author} {\bibfnamefont
  {A.}~\bibnamefont {Lichtenstein}}, \bibinfo {author} {\bibfnamefont
  {K.}~\bibnamefont {Held}}, \bibinfo {author} {\bibfnamefont {A.}~\bibnamefont
  {Katanin}}, \bibinfo {author} {\bibfnamefont {A.}~\bibnamefont {Toschi}},\
  and\ \bibinfo {author} {\bibfnamefont {G.}~\bibnamefont {Sangiovanni}},\
  }\bibfield  {title} {\bibinfo {title} {Local magnetic moments in iron and
  nickel at ambient and earth's core conditions},\ }\bibfield  {journal}
  {\bibinfo  {journal} {Nature Communications}\ }\textbf {\bibinfo {volume}
  {8}},\ \href {https://doi.org/10.1038/ncomms16062} {10.1038/ncomms16062}
  (\bibinfo {year} {2017})\BibitemShut {NoStop}%
\bibitem [{\citenamefont {Kresse}\ and\ \citenamefont
  {Furthm\"uller}(1996)}]{kresse1996}%
  \BibitemOpen
  \bibfield  {author} {\bibinfo {author} {\bibfnamefont {G.}~\bibnamefont
  {Kresse}}\ and\ \bibinfo {author} {\bibfnamefont {J.}~\bibnamefont
  {Furthm\"uller}},\ }\bibfield  {title} {\bibinfo {title} {Efficient iterative
  schemes for ab initio total-energy calculations using a plane-wave basis
  set},\ }\href {https://doi.org/10.1103/PhysRevB.54.11169} {\bibfield
  {journal} {\bibinfo  {journal} {Phys. Rev. B}\ }\textbf {\bibinfo {volume}
  {54}},\ \bibinfo {pages} {11169} (\bibinfo {year} {1996})}\BibitemShut
  {NoStop}%
\bibitem [{\citenamefont {Kresse}\ and\ \citenamefont
  {Joubert}(1999)}]{kresse1999}%
  \BibitemOpen
  \bibfield  {author} {\bibinfo {author} {\bibfnamefont {G.}~\bibnamefont
  {Kresse}}\ and\ \bibinfo {author} {\bibfnamefont {D.}~\bibnamefont
  {Joubert}},\ }\bibfield  {title} {\bibinfo {title} {From ultrasoft
  pseudopotentials to the projector augmented-wave method},\ }\href
  {https://doi.org/10.1103/PhysRevB.59.1758} {\bibfield  {journal} {\bibinfo
  {journal} {Phys. Rev. B}\ }\textbf {\bibinfo {volume} {59}},\ \bibinfo
  {pages} {1758} (\bibinfo {year} {1999})}\BibitemShut {NoStop}%
\bibitem [{\citenamefont {Bl\"ochl}(1994)}]{blochl1994}%
  \BibitemOpen
  \bibfield  {author} {\bibinfo {author} {\bibfnamefont {P.~E.}\ \bibnamefont
  {Bl\"ochl}},\ }\bibfield  {title} {\bibinfo {title} {Projector augmented-wave
  method},\ }\href {https://doi.org/10.1103/PhysRevB.50.17953} {\bibfield
  {journal} {\bibinfo  {journal} {Phys. Rev. B}\ }\textbf {\bibinfo {volume}
  {50}},\ \bibinfo {pages} {17953} (\bibinfo {year} {1994})}\BibitemShut
  {NoStop}%
\bibitem [{\citenamefont {Nos{\'{e}}}(1984)}]{Nose1984}%
  \BibitemOpen
  \bibfield  {author} {\bibinfo {author} {\bibfnamefont {S.}~\bibnamefont
  {Nos{\'{e}}}},\ }\bibfield  {title} {\bibinfo {title} {A molecular dynamics
  method for simulations in the canonical ensemble},\ }\href
  {https://doi.org/10.1080/00268978400101201} {\bibfield  {journal} {\bibinfo
  {journal} {Molecular Physics}\ }\textbf {\bibinfo {volume} {52}},\ \bibinfo
  {pages} {255} (\bibinfo {year} {1984})}\BibitemShut {NoStop}%
\bibitem [{\citenamefont {Desjarlais}\ \emph {et~al.}(2002)\citenamefont
  {Desjarlais}, \citenamefont {Kress},\ and\ \citenamefont
  {Collins}}]{Desjarlais2002}%
  \BibitemOpen
  \bibfield  {author} {\bibinfo {author} {\bibfnamefont {M.~P.}\ \bibnamefont
  {Desjarlais}}, \bibinfo {author} {\bibfnamefont {J.~D.}\ \bibnamefont
  {Kress}},\ and\ \bibinfo {author} {\bibfnamefont {L.~A.}\ \bibnamefont
  {Collins}},\ }\bibfield  {title} {\bibinfo {title} {Electrical conductivity
  for warm, dense aluminum plasmas and liquids},\ }\href
  {https://doi.org/10.1103/PhysRevE.66.025401} {\bibfield  {journal} {\bibinfo
  {journal} {Phys. Rev. E}\ }\textbf {\bibinfo {volume} {66}},\ \bibinfo
  {pages} {025401} (\bibinfo {year} {2002})}\BibitemShut {NoStop}%
\bibitem [{Wie()}]{Wien2K}%
  \BibitemOpen
  \href@noop {} {}\bibinfo {note} {P. Blaha, K. Schwarz, G. Madsen, D.
  Kvasnicka, and J. Luitz, WIEN2k, An augmented Plane Wave + Local Orbitals
  Program for Calculating Crystal Properties (Techn. Universitat Wien, Austria,
  2001).}\BibitemShut {Stop}%
\bibitem [{\citenamefont {Blaha}\ \emph {et~al.}(2020)\citenamefont {Blaha},
  \citenamefont {Schwarz}, \citenamefont {Tran}, \citenamefont {Laskowski},
  \citenamefont {Madsen},\ and\ \citenamefont {Marks}}]{Blaha2020}%
  \BibitemOpen
  \bibfield  {author} {\bibinfo {author} {\bibfnamefont {P.}~\bibnamefont
  {Blaha}}, \bibinfo {author} {\bibfnamefont {K.}~\bibnamefont {Schwarz}},
  \bibinfo {author} {\bibfnamefont {F.}~\bibnamefont {Tran}}, \bibinfo {author}
  {\bibfnamefont {R.}~\bibnamefont {Laskowski}}, \bibinfo {author}
  {\bibfnamefont {G.~K.~H.}\ \bibnamefont {Madsen}},\ and\ \bibinfo {author}
  {\bibfnamefont {L.~D.}\ \bibnamefont {Marks}},\ }\bibfield  {title} {\bibinfo
  {title} {Wien2k: An apw+lo program for calculating the properties of
  solids},\ }\href {https://doi.org/10.1063/1.5143061} {\bibfield  {journal}
  {\bibinfo  {journal} {The Journal of Chemical Physics}\ }\textbf {\bibinfo
  {volume} {152}},\ \bibinfo {pages} {074101} (\bibinfo {year} {2020})},\
  \Eprint {https://arxiv.org/abs/https://doi.org/10.1063/1.5143061}
  {https://doi.org/10.1063/1.5143061} \BibitemShut {NoStop}%
\bibitem [{\citenamefont {Parcollet}\ \emph {et~al.}(2015)\citenamefont
  {Parcollet}, \citenamefont {Ferrero}, \citenamefont {Ayral}, \citenamefont
  {Hafermann}, \citenamefont {Krivenko}, \citenamefont {Messio},\ and\
  \citenamefont {Seth}}]{triqs}%
  \BibitemOpen
  \bibfield  {author} {\bibinfo {author} {\bibfnamefont {O.}~\bibnamefont
  {Parcollet}}, \bibinfo {author} {\bibfnamefont {M.}~\bibnamefont {Ferrero}},
  \bibinfo {author} {\bibfnamefont {T.}~\bibnamefont {Ayral}}, \bibinfo
  {author} {\bibfnamefont {H.}~\bibnamefont {Hafermann}}, \bibinfo {author}
  {\bibfnamefont {I.}~\bibnamefont {Krivenko}}, \bibinfo {author}
  {\bibfnamefont {L.}~\bibnamefont {Messio}},\ and\ \bibinfo {author}
  {\bibfnamefont {P.}~\bibnamefont {Seth}},\ }\bibfield  {title} {\bibinfo
  {title} {Triqs: A toolbox for research on interacting quantum systems},\
  }\href {https://doi.org/http://dx.doi.org/10.1016/j.cpc.2015.04.023}
  {\bibfield  {journal} {\bibinfo  {journal} {Computer Physics Communications}\
  }\textbf {\bibinfo {volume} {196}},\ \bibinfo {pages} {398 } (\bibinfo {year}
  {2015})}\BibitemShut {NoStop}%
\bibitem [{\citenamefont {Aichhorn}\ \emph {et~al.}(2009)\citenamefont
  {Aichhorn}, \citenamefont {Pourovskii}, \citenamefont {Vildosola},
  \citenamefont {Ferrero}, \citenamefont {Parcollet}, \citenamefont {Miyake},
  \citenamefont {Georges},\ and\ \citenamefont
  {Biermann}}]{triqs_wien2k_interface}%
  \BibitemOpen
  \bibfield  {author} {\bibinfo {author} {\bibfnamefont {M.}~\bibnamefont
  {Aichhorn}}, \bibinfo {author} {\bibfnamefont {L.}~\bibnamefont
  {Pourovskii}}, \bibinfo {author} {\bibfnamefont {V.}~\bibnamefont
  {Vildosola}}, \bibinfo {author} {\bibfnamefont {M.}~\bibnamefont {Ferrero}},
  \bibinfo {author} {\bibfnamefont {O.}~\bibnamefont {Parcollet}}, \bibinfo
  {author} {\bibfnamefont {T.}~\bibnamefont {Miyake}}, \bibinfo {author}
  {\bibfnamefont {A.~o.}\ \bibnamefont {Georges}},\ and\ \bibinfo {author}
  {\bibfnamefont {S.}~\bibnamefont {Biermann}},\ }\bibfield  {title} {\bibinfo
  {title} {Dynamical mean-field theory within an augmented plane-wave
  framework: Assessing electronic correlations in the iron pnictide lafeaso},\
  }\href {https://doi.org/10.1103/PhysRevB.80.085101} {\bibfield  {journal}
  {\bibinfo  {journal} {Phys. Rev. B}\ }\textbf {\bibinfo {volume} {80}},\
  \bibinfo {pages} {085101} (\bibinfo {year} {2009})}\BibitemShut {NoStop}%
\bibitem [{\citenamefont {Aichhorn}\ \emph {et~al.}(2011)\citenamefont
  {Aichhorn}, \citenamefont {Pourovskii},\ and\ \citenamefont
  {Georges}}]{triqs_wien2k_full_charge_SC}%
  \BibitemOpen
  \bibfield  {author} {\bibinfo {author} {\bibfnamefont {M.}~\bibnamefont
  {Aichhorn}}, \bibinfo {author} {\bibfnamefont {L.}~\bibnamefont
  {Pourovskii}},\ and\ \bibinfo {author} {\bibfnamefont {A.}~\bibnamefont
  {Georges}},\ }\bibfield  {title} {\bibinfo {title} {Importance of electronic
  correlations for structural and magnetic properties of the iron pnictide
  superconductor lafeaso},\ }\href {https://doi.org/10.1103/PhysRevB.84.054529}
  {\bibfield  {journal} {\bibinfo  {journal} {Phys. Rev. B}\ }\textbf {\bibinfo
  {volume} {84}},\ \bibinfo {pages} {054529} (\bibinfo {year}
  {2011})}\BibitemShut {NoStop}%
\bibitem [{\citenamefont {Aichhorn}\ \emph {et~al.}(2016)\citenamefont
  {Aichhorn}, \citenamefont {Pourovskii}, \citenamefont {Seth}, \citenamefont
  {Vildosola}, \citenamefont {Zingl}, \citenamefont {Peil}, \citenamefont
  {Deng}, \citenamefont {Mravlje}, \citenamefont {Kraberger}, \citenamefont
  {Martins}, \citenamefont {Ferrero},\ and\ \citenamefont
  {Parcollet}}]{TRIQS/DFTTools}%
  \BibitemOpen
  \bibfield  {author} {\bibinfo {author} {\bibfnamefont {M.}~\bibnamefont
  {Aichhorn}}, \bibinfo {author} {\bibfnamefont {L.}~\bibnamefont
  {Pourovskii}}, \bibinfo {author} {\bibfnamefont {P.}~\bibnamefont {Seth}},
  \bibinfo {author} {\bibfnamefont {V.}~\bibnamefont {Vildosola}}, \bibinfo
  {author} {\bibfnamefont {M.}~\bibnamefont {Zingl}}, \bibinfo {author}
  {\bibfnamefont {O.~E.}\ \bibnamefont {Peil}}, \bibinfo {author}
  {\bibfnamefont {X.}~\bibnamefont {Deng}}, \bibinfo {author} {\bibfnamefont
  {J.}~\bibnamefont {Mravlje}}, \bibinfo {author} {\bibfnamefont {G.~J.}\
  \bibnamefont {Kraberger}}, \bibinfo {author} {\bibfnamefont {C.}~\bibnamefont
  {Martins}}, \bibinfo {author} {\bibfnamefont {M.}~\bibnamefont {Ferrero}},\
  and\ \bibinfo {author} {\bibfnamefont {O.}~\bibnamefont {Parcollet}},\
  }\bibfield  {title} {\bibinfo {title} {Triqs/dfttools: A \{TRIQS\}
  application for ab initio calculations of correlated materials},\ }\href
  {https://doi.org/http://dx.doi.org/10.1016/j.cpc.2016.03.014} {\bibfield
  {journal} {\bibinfo  {journal} {Computer Physics Communications}\ }\textbf
  {\bibinfo {volume} {204}},\ \bibinfo {pages} {200 } (\bibinfo {year}
  {2016})}\BibitemShut {NoStop}%
\bibitem [{\citenamefont {Werner}\ \emph {et~al.}(2006)\citenamefont {Werner},
  \citenamefont {Comanac}, \citenamefont {de' Medici}, \citenamefont {Troyer},\
  and\ \citenamefont {Millis}}]{ctseg1}%
  \BibitemOpen
  \bibfield  {author} {\bibinfo {author} {\bibfnamefont {P.}~\bibnamefont
  {Werner}}, \bibinfo {author} {\bibfnamefont {A.}~\bibnamefont {Comanac}},
  \bibinfo {author} {\bibfnamefont {L.}~\bibnamefont {de' Medici}}, \bibinfo
  {author} {\bibfnamefont {M.}~\bibnamefont {Troyer}},\ and\ \bibinfo {author}
  {\bibfnamefont {A.~J.}\ \bibnamefont {Millis}},\ }\bibfield  {title}
  {\bibinfo {title} {Continuous-time solver for quantum impurity models},\
  }\href {https://doi.org/10.1103/PhysRevLett.97.076405} {\bibfield  {journal}
  {\bibinfo  {journal} {Phys. Rev. Lett.}\ }\textbf {\bibinfo {volume} {97}},\
  \bibinfo {pages} {076405} (\bibinfo {year} {2006})}\BibitemShut {NoStop}%
\bibitem [{\citenamefont {Werner}\ and\ \citenamefont {Millis}(2006)}]{ctseg2}%
  \BibitemOpen
  \bibfield  {author} {\bibinfo {author} {\bibfnamefont {P.}~\bibnamefont
  {Werner}}\ and\ \bibinfo {author} {\bibfnamefont {A.~J.}\ \bibnamefont
  {Millis}},\ }\bibfield  {title} {\bibinfo {title} {Hybridization expansion
  impurity solver: General formulation and application to kondo lattice and
  two-orbital models},\ }\href {https://doi.org/10.1103/PhysRevB.74.155107}
  {\bibfield  {journal} {\bibinfo  {journal} {Phys. Rev. B}\ }\textbf {\bibinfo
  {volume} {74}},\ \bibinfo {pages} {155107} (\bibinfo {year}
  {2006})}\BibitemShut {NoStop}%
\bibitem [{\citenamefont {Kotliar}\ \emph {et~al.}(2006)\citenamefont
  {Kotliar}, \citenamefont {Savrasov}, \citenamefont {Haule}, \citenamefont
  {Oudovenko}, \citenamefont {Parcollet},\ and\ \citenamefont
  {Marianetti}}]{kotliar_elec_struc_dmft_2006}%
  \BibitemOpen
  \bibfield  {author} {\bibinfo {author} {\bibfnamefont {G.}~\bibnamefont
  {Kotliar}}, \bibinfo {author} {\bibfnamefont {S.~Y.}\ \bibnamefont
  {Savrasov}}, \bibinfo {author} {\bibfnamefont {K.}~\bibnamefont {Haule}},
  \bibinfo {author} {\bibfnamefont {V.~S.}\ \bibnamefont {Oudovenko}}, \bibinfo
  {author} {\bibfnamefont {O.}~\bibnamefont {Parcollet}},\ and\ \bibinfo
  {author} {\bibfnamefont {C.}~\bibnamefont {Marianetti}},\ }\bibfield  {title}
  {\bibinfo {title} {Electronic structure calculations with dynamical
  mean-field theory},\ }\href@noop {} {\bibfield  {journal} {\bibinfo
  {journal} {Reviews of Modern Physics}\ }\textbf {\bibinfo {volume} {78}},\
  \bibinfo {pages} {865} (\bibinfo {year} {2006})}\BibitemShut {NoStop}%
\bibitem [{\citenamefont {Driscoll}\ and\ \citenamefont
  {Bercovici}(2014)}]{Driscoll2014}%
  \BibitemOpen
  \bibfield  {author} {\bibinfo {author} {\bibfnamefont {P.}~\bibnamefont
  {Driscoll}}\ and\ \bibinfo {author} {\bibfnamefont {D.}~\bibnamefont
  {Bercovici}},\ }\bibfield  {title} {\bibinfo {title} {On the thermal and
  magnetic histories of earth and venus: Influences of melting, radioactivity,
  and conductivity},\ }\href {https://doi.org/10.1016/j.pepi.2014.08.004}
  {\bibfield  {journal} {\bibinfo  {journal} {Physics of the Earth and
  Planetary Interiors}\ }\textbf {\bibinfo {volume} {236}},\ \bibinfo {pages}
  {36} (\bibinfo {year} {2014})}\BibitemShut {NoStop}%
\bibitem [{\citenamefont {Landeau}\ \emph {et~al.}(2022)\citenamefont
  {Landeau}, \citenamefont {Fournier}, \citenamefont {Nataf}, \citenamefont
  {C{\'e}bron},\ and\ \citenamefont {Schaeffer}}]{Landeau2022}%
  \BibitemOpen
  \bibfield  {author} {\bibinfo {author} {\bibfnamefont {M.}~\bibnamefont
  {Landeau}}, \bibinfo {author} {\bibfnamefont {A.}~\bibnamefont {Fournier}},
  \bibinfo {author} {\bibfnamefont {H.-C.}\ \bibnamefont {Nataf}}, \bibinfo
  {author} {\bibfnamefont {D.}~\bibnamefont {C{\'e}bron}},\ and\ \bibinfo
  {author} {\bibfnamefont {N.}~\bibnamefont {Schaeffer}},\ }\bibfield  {title}
  {\bibinfo {title} {Sustaining earth's magnetic dynamo},\ }\href
  {https://doi.org/10.1038/s43017-022-00264-1} {\bibfield  {journal} {\bibinfo
  {journal} {Nature Reviews Earth {\&} Environment}\ }\textbf {\bibinfo
  {volume} {3}},\ \bibinfo {pages} {255} (\bibinfo {year} {2022})}\BibitemShut
  {NoStop}%
\bibitem [{\citenamefont {Dreibus}\ and\ \citenamefont
  {Palme}(1996)}]{Dreibus1996}%
  \BibitemOpen
  \bibfield  {author} {\bibinfo {author} {\bibfnamefont {G.}~\bibnamefont
  {Dreibus}}\ and\ \bibinfo {author} {\bibfnamefont {H.}~\bibnamefont
  {Palme}},\ }\bibfield  {title} {\bibinfo {title} {Cosmochemical constraints
  on the sulfur content in the earth{\textquotesingle}s core},\ }\href
  {https://doi.org/10.1016/0016-7037(96)00028-2} {\bibfield  {journal}
  {\bibinfo  {journal} {Geochimica et Cosmochimica Acta}\ }\textbf {\bibinfo
  {volume} {60}},\ \bibinfo {pages} {1125} (\bibinfo {year}
  {1996})}\BibitemShut {NoStop}%
\bibitem [{\citenamefont {Namur}\ \emph {et~al.}(2016)\citenamefont {Namur},
  \citenamefont {Charlier}, \citenamefont {Holtz}, \citenamefont {Cartier},\
  and\ \citenamefont {McCammon}}]{Namur2016}%
  \BibitemOpen
  \bibfield  {author} {\bibinfo {author} {\bibfnamefont {O.}~\bibnamefont
  {Namur}}, \bibinfo {author} {\bibfnamefont {B.}~\bibnamefont {Charlier}},
  \bibinfo {author} {\bibfnamefont {F.}~\bibnamefont {Holtz}}, \bibinfo
  {author} {\bibfnamefont {C.}~\bibnamefont {Cartier}},\ and\ \bibinfo {author}
  {\bibfnamefont {C.}~\bibnamefont {McCammon}},\ }\bibfield  {title} {\bibinfo
  {title} {Sulfur solubility in reduced mafic silicate melts: Implications for
  the speciation and distribution of sulfur on mercury},\ }\href
  {https://doi.org/10.1016/j.epsl.2016.05.024} {\bibfield  {journal} {\bibinfo
  {journal} {Earth and Planetary Science Letters}\ }\textbf {\bibinfo {volume}
  {448}},\ \bibinfo {pages} {102} (\bibinfo {year} {2016})}\BibitemShut
  {NoStop}%
\bibitem [{\citenamefont {Ohta}\ \emph {et~al.}(2012)\citenamefont {Ohta},
  \citenamefont {Cohen}, \citenamefont {Hirose}, \citenamefont {Haule},
  \citenamefont {Shimizu},\ and\ \citenamefont {Ohishi}}]{ohta2012}%
  \BibitemOpen
  \bibfield  {author} {\bibinfo {author} {\bibfnamefont {K.}~\bibnamefont
  {Ohta}}, \bibinfo {author} {\bibfnamefont {R.~E.}\ \bibnamefont {Cohen}},
  \bibinfo {author} {\bibfnamefont {K.}~\bibnamefont {Hirose}}, \bibinfo
  {author} {\bibfnamefont {K.}~\bibnamefont {Haule}}, \bibinfo {author}
  {\bibfnamefont {K.}~\bibnamefont {Shimizu}},\ and\ \bibinfo {author}
  {\bibfnamefont {Y.}~\bibnamefont {Ohishi}},\ }\bibfield  {title} {\bibinfo
  {title} {Experimental and theoretical evidence for pressure-induced
  metallization in feo with rocksalt-type structure},\ }\href
  {https://doi.org/10.1103/PhysRevLett.108.026403} {\bibfield  {journal}
  {\bibinfo  {journal} {Phys. Rev. Lett.}\ }\textbf {\bibinfo {volume} {108}},\
  \bibinfo {pages} {026403} (\bibinfo {year} {2012})}\BibitemShut {NoStop}%
\bibitem [{\citenamefont {Leonov}\ \emph {et~al.}(2017)\citenamefont {Leonov},
  \citenamefont {Ponomareva}, \citenamefont {Nazarov},\ and\ \citenamefont
  {Abrikosov}}]{leonov2017}%
  \BibitemOpen
  \bibfield  {author} {\bibinfo {author} {\bibfnamefont {I.}~\bibnamefont
  {Leonov}}, \bibinfo {author} {\bibfnamefont {A.~V.}\ \bibnamefont
  {Ponomareva}}, \bibinfo {author} {\bibfnamefont {R.}~\bibnamefont
  {Nazarov}},\ and\ \bibinfo {author} {\bibfnamefont {I.~A.}\ \bibnamefont
  {Abrikosov}},\ }\bibfield  {title} {\bibinfo {title} {Pressure-induced
  spin-state transition of iron in magnesiow\"ustite (fe,mg)o},\ }\href
  {https://doi.org/10.1103/PhysRevB.96.075136} {\bibfield  {journal} {\bibinfo
  {journal} {Phys. Rev. B}\ }\textbf {\bibinfo {volume} {96}},\ \bibinfo
  {pages} {075136} (\bibinfo {year} {2017})}\BibitemShut {NoStop}%
\bibitem [{\citenamefont {Leonov}\ \emph {et~al.}(2020)\citenamefont {Leonov},
  \citenamefont {Shorikov}, \citenamefont {Anisimov},\ and\ \citenamefont
  {Abrikosov}}]{leonov2020}%
  \BibitemOpen
  \bibfield  {author} {\bibinfo {author} {\bibfnamefont {I.}~\bibnamefont
  {Leonov}}, \bibinfo {author} {\bibfnamefont {A.~O.}\ \bibnamefont
  {Shorikov}}, \bibinfo {author} {\bibfnamefont {V.~I.}\ \bibnamefont
  {Anisimov}},\ and\ \bibinfo {author} {\bibfnamefont {I.~A.}\ \bibnamefont
  {Abrikosov}},\ }\bibfield  {title} {\bibinfo {title} {Emergence of quantum
  critical charge and spin-state fluctuations near the pressure-induced mott
  transition in mno, feo, coo, and nio},\ }\href
  {https://doi.org/10.1103/PhysRevB.101.245144} {\bibfield  {journal} {\bibinfo
   {journal} {Phys. Rev. B}\ }\textbf {\bibinfo {volume} {101}},\ \bibinfo
  {pages} {245144} (\bibinfo {year} {2020})}\BibitemShut {NoStop}%
\bibitem [{\citenamefont {Ho}\ \emph {et~al.}()\citenamefont {Ho},
  \citenamefont {Zhang}, \citenamefont {Haule}, \citenamefont {Jackson},
  \citenamefont {Dobrosavljevic},\ and\ \citenamefont
  {Dobrosavljevic}}]{WaiGa2023}%
  \BibitemOpen
  \bibfield  {author} {\bibinfo {author} {\bibfnamefont {W.~D.}\ \bibnamefont
  {Ho}}, \bibinfo {author} {\bibfnamefont {P.}~\bibnamefont {Zhang}}, \bibinfo
  {author} {\bibfnamefont {K.}~\bibnamefont {Haule}}, \bibinfo {author}
  {\bibfnamefont {J.}~\bibnamefont {Jackson}}, \bibinfo {author} {\bibfnamefont
  {V.}~\bibnamefont {Dobrosavljevic}},\ and\ \bibinfo {author} {\bibfnamefont
  {V.}~\bibnamefont {Dobrosavljevic}},\ }\bibfield  {title} {\bibinfo {title}
  {{Quantum critical phase of FeO spans conditions of Earth's lower mantle}},\
  }\bibinfo {note} {arXiv:2301.047771}\BibitemShut {NoStop}%
\end{thebibliography}
\providecommand{\noopsort}[1]{}\providecommand{\singleletter}[1]{#1}%

\end{document}